\documentclass[pra,amsmath,amssymb,twocolumn,superscriptaddress]{revtex4-2}
\usepackage{amsmath}
\usepackage{amssymb}
\usepackage{amstext}
\usepackage{amsfonts}
\usepackage{amsxtra}
\usepackage{bm}
\usepackage[usenames]{color}
\usepackage{grffile}
\usepackage{soul}
\usepackage{epstopdf}
\usepackage{xcolor}
\usepackage{graphicx}
\usepackage{marvosym}
\usepackage{wasysym}
\usepackage{dsfont}

\usepackage{hyperref}
\hypersetup{
	colorlinks=true,
	linkcolor=blue,
	citecolor=blue,
	urlcolor=blue}

{%
	\setlength{\fboxsep}{0pt}%
	\setlength{\fboxrule}{1pt}%
}%
\makeatletter
\let\cat@comma@active\@empty
\makeatother

\begin{document}
	

	\title{Kibble-Zurek scaling of the superfluid-supersolid transition in an elongated dipolar gas}
	
	\author{Wyatt Kirkby}
	\affiliation{Kirchhoff-Institut f\"ur Physik,
		Universit\"at Heidelberg,
		Im~Neuenheimer~Feld~227,
		69120~Heidelberg, Germany}
	\affiliation{Physikalisches Institut,
		Universit\"at Heidelberg,
		Im~Neuenheimer~Feld~226,
		69120~Heidelberg, Germany}

    \author{Hayder Salman}
	\affiliation{School of Engineering, Mathematics, and Physics, University~of~East~Anglia, Norwich~Research~Park, Norwich, NR4 7TJ, UK}
	\affiliation{Centre for Photonics and Quantum Science, University~of~East~Anglia, Norwich~Research~Park, Norwich, NR4 7TJ, UK}
    
    \author{Thomas Gasenzer}
	\affiliation{Kirchhoff-Institut f\"ur Physik,
		Universit\"at Heidelberg,
		Im~Neuenheimer~Feld~227,
		69120~Heidelberg, Germany}
	\affiliation{Institut f\"ur Theoretische Physik,
		Universit\"at Heidelberg,
		Philosophenweg 16,
		69120~Heidelberg, Germany}
        
    \author{Lauriane Chomaz}
	\affiliation{Physikalisches Institut,
		Universit\"at Heidelberg,
		Im~Neuenheimer~Feld~226,
		69120~Heidelberg, Germany}
	\begin{abstract}
		We simulate interaction quenches crossing from a superfluid to a supersolid state in a dipolar quantum gas of ${}^{164}\mathrm{Dy}$ atoms, trapped in an elongated tube with periodic boundary conditions, via the extended Gross-Pitaevskii equation. 
		A freeze-out time is observed through a delay in supersolid formation after crossing the critical point. We compute the density-density correlations at the freeze-out time and extract the frozen correlation length for the solid order. An analysis of the freeze-out time and correlation length versus the interaction quench rate allows us to extract universal exponents corresponding to the relaxation time and correlation length based on predictions of the Kibble-Zurek mechanism. Over several orders of magnitude, clear power-law scaling is observed for both the freeze-out time and the correlation length, and the corresponding exponents are compatible with predictions based on the excitation spectrum calculated via Bogoliubov theory. Defects due to independent local breaking of translational symmetry, contributing to globally incommensurate supersolid order, are identified, and their number at the freeze-out time is found to also scale as a power law. Our results support the hypothesis of a continuous transition whose universality class remains to be determined but appears to differ from that of the (1+1)D XY model.
	\end{abstract}
	\date{\today}
	
	\maketitle
	
	\section{Introduction}
	\label{sec:intro}

 	Spontaneous breaking of translational symmetry in a superfluid can result in an exotic phase of matter that maintains phase coherence while exhibiting periodic crystalline order: a supersolid \cite{Penrose1956,gross1957unified,gross1958classical,andreev1969quantum,leggett1970can,boninsegni2012colloquium}. Supersolids have been realized in ultracold quantum gases using spin-orbit-coupled systems \cite{li2017stripe,putra2020spatial},
	and with strongly-magnetic dipolar atoms \cite{Tanzi2019,Bottcher2019,Chomaz2019}. In the latter, the transition to a solid phase can be achieved either starting from a thermal gas and quenching the temperature \cite{Chomaz2019,Sohmen2021,Bland2022Two}, or starting from a superfluid and changing the interparticle interaction strength \cite{Tanzi2019,Bottcher2019,Chomaz2019}. Furthermore, depending on the system parameters and geometry, the quantum phase transition from superfluid to supersolid has been found to be either continuous or discontinuous \cite{Roccuzzo2019,blakie2020supersolidity,Smith2023supersolidity,zhang2019supersolidity,zhang2021phases,Ripley2023Two,Zhang2024}, however the exact critical behavior and the corresponding universality classes of these transitions have yet to be determined.

	\begin{figure}[!ht]
		\centering
		\includegraphics[width=1\columnwidth]{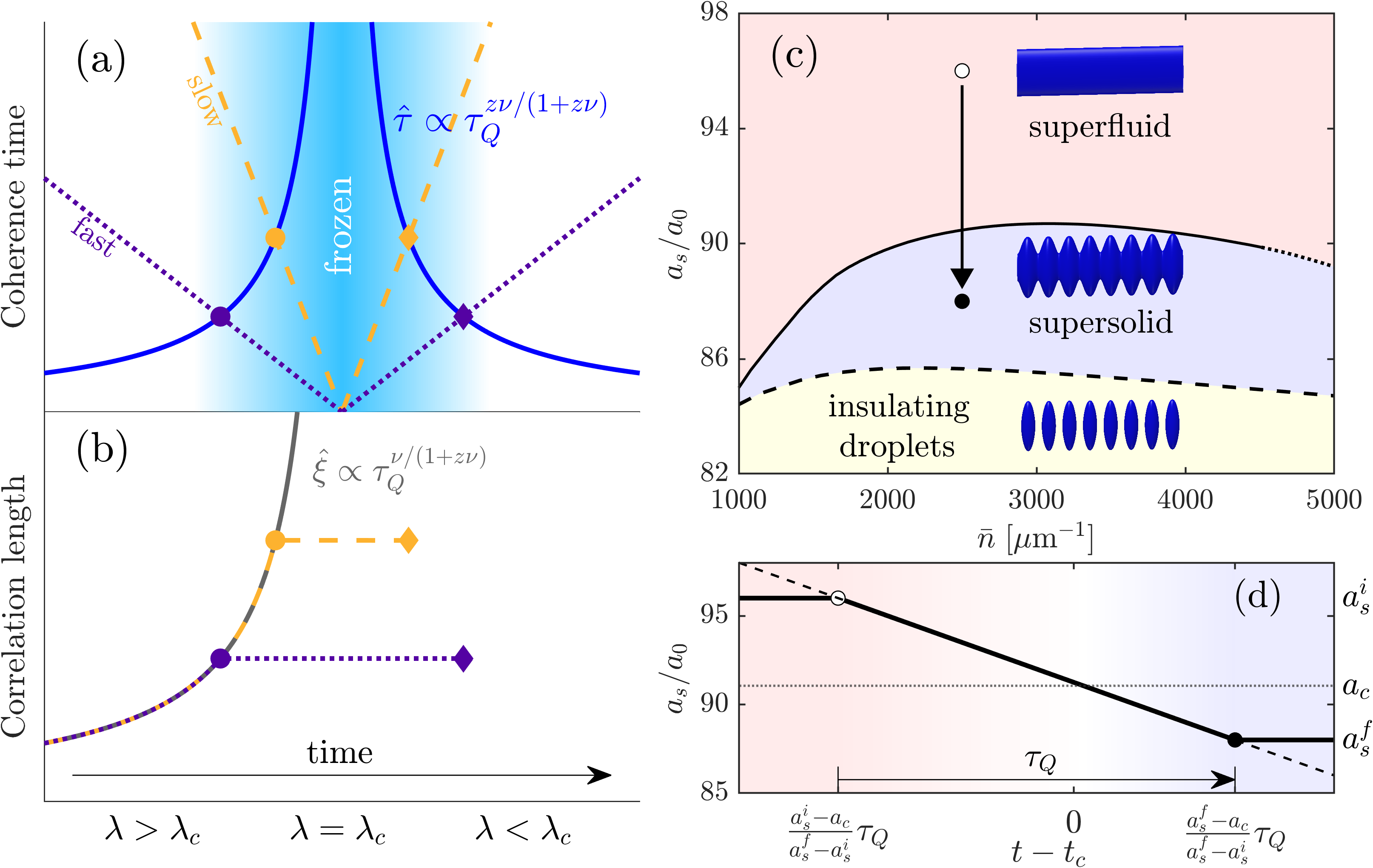}
		\caption{KZM in quenches across the superfluid-supersolid transition. Panel (a) depicts the typical KZM schematic. The coherence time $\tau$ (blue curve) diverges at the critical point. Two quench ramps linear in time $t$ (slow: green-dashed, fast: yellow-dotted) of a control parameter are indicated. The freeze-out time $\hat{\tau}$ occurs when $|t|$ crosses $\tau$. Panel (b) depicts the diverging correlation length $\xi$ (grey solid curve), which becomes frozen at $\hat{\xi}$ as depicted here for the two ramps of panel (a). Panel (c) shows the phase diagram for the elongated dipolar system, with a continuous (solid line) or discontinuous (dotted line) transition from superfluid to supersolid phases, and a crossover from supersolid to insulating droplet (dashed line). Sample isodensity surfaces for each phase are shown. The start (empty circle) and end points (filled circle) for the quenches considered in this paper are indicated. The quench protocol is shown in (d), with the total quench time $\tau_Q$ varying throughout the paper. Most quenches follow the solid line, while for extremely fast quenches $(\tau_Q< 10\,\mathrm{ms})$, the dashed line is followed.}
		\label{fig:kzfig}
	\end{figure}
 
	When tuning a control parameter $\lambda$ towards its critical value $\lambda_{\mathrm{c}}$ at a continuous phase transition, characteristic scales diverge following universal power laws, defining the critical behavior at the phase transition and being independent of the microscopic details. Precisely, the correlation length $\xi$ and relaxation time $\tau$  diverge as $\xi(\lambda)\sim|\lambda-\lambda_{\mathrm{c}}|^{-\nu}$ and $\tau(\lambda)\sim \xi^{z} \sim|\lambda-\lambda_{\mathrm{c}}|^{-\nu z}$, respectively, with $\nu$ and $z$ the critical exponents determined by the universality class of the transition \cite{Fisher1968,Odor2004,ZinnJustin2002Book}, see Fig.~\ref{fig:kzfig} (a) and (b).

  The theory of the Kibble-Zurek mechanism (KZM) \cite{Kibble1976,Kibble1980,Zurek1985,Zurek1996} provides a framework for understanding symmetry breaking in a dynamical setting. It predicts that critical slowing down and the divergence of both the characteristic length and time scales prevent the system from preserving an instantaneous parametric equilibrium state in quenches across continuous phase transitions. More specifically, in a quench of the control parameter $\lambda$, such as $\lambda-\lambda_{\mathrm{c}}\propto t/\tau_Q$, the relaxation time $\tau(t)$ rapidly increases and surpasses the time left to the transition at time $t=-\hat{\tau}=-\tau(\hat{\tau})$, so-called freezing-out time. For $t>-\hat{\tau}$, adiabaticity is broken, the system becomes frozen, and fluctuations present remain embedded in the system until it unfreezes, which similarly occurs at a certain time after crossing the critical point, namely for $t>\hat{\tau}$. During the freeze-out period and as adiabaticity is lost, domains with distinct local order roughly of size $\hat{\xi}=\xi(\hat{\tau})$ persist.

 	Based on the equilibrium scaling laws above, the freeze-out time and domain size for a linear quench of the order parameter at a constant rate $\tau_Q$ scale as,
	\begin{align}
		\hat{\tau}\propto \tau_Q^{z\nu/(1+z\nu)}\equiv\tau_Q^{\zeta_\text{KZ}} \;
  		\quad
        \text{and}\quad
		\hat{\xi}\propto \tau_Q^{\nu/(1+z\nu)}\equiv\tau_Q^{\nu_\text{KZ}}\;,
        \label{eq:kzscaling}
	\end{align}
	respectively, see Fig.~\ref{fig:kzfig} (a) and (b). Higher quench rates lead to the breakdown of adiabaticity at lower values of $\hat{\tau}$ (faster supersolid formation) and correspondingly lower $\hat{\xi}$ (smaller domain sizes). The boundaries of these domains can be identified as defects, corresponding to sudden changes in the value of the order parameter in space, and separating independent causally-disconnected regions where the symmetry is independently broken. By studying the scaling of the freeze-out time, domain size, or remaining defects, one can extract critical scaling exponents of the underlying transition and thereby probe the transition universality class. 	

	The KZM has been successfully applied to a wide range of systems~\cite{delcampo2014,kibble2007,Zurek1996}. This is in particular true for quantum gases where it has been studied both theoretically \cite{Uhlmann2007,Zurek2009,Damski2010,delcampo2011,Das2012,delcampo2013,Su2013,Beugnon2017,Liu2019,Comaron2019,Jiang2019,Liu2020,Thudiyangal2024,Proukakis2024Book,Wheeler2025}, and experimentally, via temperature quenches \cite{Weiler2008,Lamporesi2013,Corman2014,Navon2015,Chomaz2015,Donadello2016,Clark2016,liu2018dynamical,Ko2019,Goo2021,Liu2021,Goo2022,Rabga2023,Allman2024} and in quantum phase transitions \cite{Zurek2005,Polkovnikov2005,Sadler2006,Chen2011,Clark2016,Anquez2016,Feng2017,Chen2019,Yi2020,Lee2024}. The transition to supersolid states offers a new context, where the KZM could be induced and its study may yield new insights into the nature of the underlying transitions.
		
	In this paper, we study the KZM in the context of a continuous quantum phase transition from a superfluid to a one-dimensional supersolid in an elongated tube at a fixed density by tuning the $s$-wave scattering length $a_s$, see Fig.~\ref{fig:kzfig} (c). We cross the transition at a finite rate, and examine the scalings of the characteristic scales with the quench rate. In Section \ref{sec:model}, we present the equations of motion for the dipolar gas, illustrate how the ground state phase diagram is calculated, and present details of our dynamical simualtions. Sections \ref{subsec:scaling}-\ref{subsec:defects} presents our observations for power-law scalings of the freeze-out time, domain size, and defect densities, respectively, along with the corresponding scaling exponents. In Sec. \ref{subsec:scaling}, based on these KZM scalings, we extract the critical exponents $\nu$ and $z$ of the superfluid-to-supersolid transition and comment on its universality class.
	
	\section{Model}
	\label{sec:model}
	\subsection{Equations of motion}
	\label{subsec:system}
	
	To simulate the dipolar gas, we begin with the extended Gross-Pitaevskii equation (eGPE),
	\begin{align}
		&\mathrm{i}\hbar\frac{\partial \Psi(\textbf{r},t)}{\partial t}=\Biggl[-\frac{\hbar^2\nabla^2}{2m}+V(\textbf{r})
		+g|\Psi(\textbf{r},t)|^2\nonumber\\
			&
		+\gamma_\mathrm{QF}|\Psi(\textbf{r},t)|^3
	+\int\mathrm{d}^3r'U_{\text{dd}}(\textbf{r}-\textbf{r}')|\Psi(\textbf{r}',t)|^2\Biggr]\Psi(\textbf{r},t),
		\label{eq:GPE3D}
	\end{align}
	for atomic species with mass $m$ and contact interaction strength $g=4\pi \hbar^2a_s/m$ for an $s$-wave scattering length $a_s$. We assume the system is harmonically trapped along two directions via $V(\textbf{r})=\frac{1}{2}m(\omega_y^2y^2+\omega_z^2z^2)$ while the $x$-direction is untrapped. The dipole-dipole interaction (DDI) is given by 
	$U_{\mathrm{dd}}(\textbf{r})=\mu_0\mu_m^2(1-3\cos^2\theta)/(4\pi|\textbf{r}|^3)$ where $\theta$ is the angle between $\textbf{r}$ and the $z$ axis, along which the dipoles are polarized. Each atom has a magnetic moment $\mu_m$, and $\mu_0$ is the vacuum permeability. The supersolid phase is protected from collapse \cite{FerrierBarbut2016,Wachtler2016a,Chomaz2016} by the beyond-mean field quantum fluctuations term proportional to $\gamma_\mathrm{QF}=\frac{32}{3}g\sqrt{a_s^3/\pi}\mathcal{Q}_5(\epsilon_{\text{dd}})$ \cite{Schuetzhold2005,Lima2011a}, with $\mathcal{Q}_5(\epsilon) = \mathrm{Re}\big\{\int_0^1\mathrm{d}u[1 + (3u^2-1)\epsilon]^{5/2}\big\}$, where $\mathrm{Re}\{\cdot\}$ denotes the real part, $\epsilon_{\text{dd}}=a_{\text{dd}}/a_s$ and $a_{\text{dd}}=\mu_0\mu_m^2/(12\pi\hbar^2)$. The wavefunction is normalized to the total particle number $N=\int\mathrm{d}^3r\;|\Psi(\textbf{r})|^2$. Note that the eGPE does not account for the quantum fluctuations in a fully self-consistent manner (as the higher-order diagrams are taken into account non-dynamically). The resulting equation of motion Eq.~\eqref{eq:GPE3D} is thereby again mean-field like.
	
	We use a reduced 3D model introduced in Refs.\  \cite{Blakie2020variational,Pal2020,blakie2020supersolidity}, which assumes that the wavefunction $\Psi(\textbf{r})$ is separable,
	\begin{equation}
		\Psi(\textbf{r})=\varphi(y,z)\psi(x,t)\;,
	\end{equation}
	with radial wavefunction,
	\begin{equation}
		\varphi(y,z)=\frac{1}{\ell\sqrt{\pi}}\exp\left[-\frac{\eta y^2+z^2/\eta}{2\ell^2}\right]\;,
	\end{equation}
	where $\ell$ and $\eta$ are variational parameters corresponding to the $1/\mathrm{e}$ condensate width and ellipticity (due to magnetostriction), respectively, obtained via minimization of the eGPE in imaginary time. The reduced eGPE then reads,
	\begin{align}
		\mathrm{i}\hbar\frac{\partial \psi(x,t)}{\partial t}&=\Biggl[-\frac{\hbar^2}{2m}\frac{\partial^2}{\partial x^2}+\mathcal{E}_{\perp}+\frac{g}{2 \pi  \ell ^2}|\psi(x,t)|^2
		\nonumber\\
		&+\frac{2\gamma_\mathrm{QF}}{5 \pi ^{3/2} \ell ^3}|\psi(x,t)|^3
		\nonumber\\
		&+\int\mathrm{d}x'U^{\mathrm{1D}}(x-x')|\psi(x',t)|^2\Biggr]\psi(x,t)\,,
		\label{eq:1DGPE}
	\end{align}
	where
	$\mathcal{E}_{\perp}={\hbar^2}\left(\eta+{1}/{\eta}\right)/({4m\ell^2})+m\ell^2\left({\omega_y^2}/{\eta}+\eta\omega_z^2\right)/4$, and where we have defined the effective quasi-1D DDI $U^{\mathrm{1D}}(x-x')$, which, for dipoles polarized along the $z$ axis, has the Fourier representation,
	$\tilde{U}^{1\text{D}}(k)={\mu_0\mu_m^2}\big[{4-2\eta}-{3} \sqrt{\eta}k^2\ell^2\mathrm{e}^{\sqrt{\eta}k^2\ell^2/2}\mathrm{Ei}\left(-\sqrt{\eta}k^2\ell^2/2\right)\big]/[{12\pi\ell^2}({1+\eta})]$,
	where $\mathrm{Ei}(k)$ is the exponential integral, and $k$ is the conjugate momentum to the $x$ coordinate. We will focus on an elongated gas of ${}^{164}\mathrm{Dy}$, with $\mu_m=9.93\,\mu_B$ in terms of the Bohr magneton $\mu_B$, in an elongated tube with trapping frequencies $\{\omega_y,\omega_z\}=2\pi\times \{150,150\}\,$Hz. The eGPE results in unitary time evolution, conserving total atom number. However, due to the fact that we are quenching the scattering length, total energy is not conserved over the course of the dynamics.

    \subsection{Ground states}
	\label{subsec:groundstate}
	Prior to performing any quenches, we consider the ground state properties of the system. In order to determine the ground state of the system, and therefore the phase diagram, we perform an imaginary-time evolution of the reduced eGPE given in Eq.\ \eqref{eq:1DGPE}. We apply a split-step Fourier algorithm with an adaptive time step. For the ground state and dynamic simulations (next section), a spatial grid of approximately $0.04\,\mu$m is selected, such that there are always approximately 5-8 grid points per healing length, depending on the location in the phase diagram. By evaluating the long-range interaction term in momentum space, it is computationally efficient to simulate a single unit cell of the supersolid crystal (for ground-state calculations only) and allow alias Fourier copies of the simulation cell to self-interact \cite{Blakie2023}. By minimizing the energy per particle with respect to the variational parameters $\ell$, $\eta$, and primitive unit cell length $L_{\mathrm{uc}}$, it becomes possible to determine the ground state at each point in the phase diagram.
 
    The phase diagram for this system is sketched in Fig.~\ref{fig:kzfig} (c), cf.~Refs.\ \cite{blakie2020supersolidity,Smith2023supersolidity}. In order to characterize the phases of the system, we make use of Leggett's estimate on the upper bound of the superfluid fraction \cite{leggett1970can},
	\begin{equation}
		f^s=\frac{L^2}{N}\left[\int\mathrm{d}x\left(\int\mathrm{d}y\,\mathrm{d}z|\Psi|^2\right)^{-1}\right]^{-1}\;,
		\label{eq:Leggett}
	\end{equation}
	where $L=\int\mathrm{d}x$. A value of $f^s=1$ indicates the uniform superfluid regime, $0.1<f^s<1$ for the supersolid phase, and $f^s\leq 0.1$ corresponds to insulating droplets. The boundary to the latter is not captured by our eGPE simulation and the threshold choice of 0.1 is arbitrary, yet the details of this transition are irrelevant to the present work. We will be using $f^s$ as an order parameter for the superfluid-supersolid transition. It should be remarked that the expression in Eq.~\eqref{eq:Leggett} corresponds strictly to an estimate on the \textit{upper} bound of the superfluid fraction. However, a lower bound has also been derived \cite{Leggett1998}, and in the case of a 1D system where the modulation exists along one direction (the case we consider here), the two bounds are exactly equal. Using the density modulation contrast $\mathcal{C}=[\max(n)-\min(n)]/[\max(n)+\min(n)]$ as an order parameter instead of $f^s$ was found to give similar results.
	
	The order of the transition depends on the linear density \cite{blakie2020supersolidity,Alana2022,Smith2023supersolidity} (see also the experiment in Ref. \cite{Biagioni2022}), 
	$\bar{n}=N/L$, and is discontinuous at both low ($\bar{n}\lesssim 800\,\mu \mathrm{m}^{-1}$) and high ($\bar{n}\gtrsim 4500\,\mu \mathrm{m}^{-1}$) densities. In the intermediate regime, the transition is continuous: the ground state’s modulation emerges smoothly, precisely when the roton mode—calculated with Bogoliubov theory—becomes unstable in the dispersion relation. In this way, the continuous phase transition can be pinpointed precisely, whereas in the first‑order regime a bistable region emerges where the true ground state is already modulated, even though the roton has not yet fully softened. In this paper, as a first study of the KZM applied to dipolar supersolids, we restrict our quenches to an intermediate regime in which the transition is expected to be continuous, taking $\bar{n}=2500\,\mu \mathrm{m}^{-1}$, for which the superfluid-supersolid transition is found to occur at $a_s = a_{\mathrm{c}}^{\mathrm{(GS)}}\approx90.5\,a_0$.
	
	\subsection{Ramp into the supersolid phase}
	\label{subsec:quench}
	
	In this work, the initial state is selected to be the ground state of the uniform superfluid regime at $a_s^{i}=96\,a_0$. Thermal noise at a temperature of $T=20\,$nK is then added (see Appendix \ref{app:SMnoise} for details), typically adding approximately $3$-$5\%$ excited atoms, and allowed to equilibrate for $10\,$ms. The initial noise is crucial to trigger the dynamics away from the unstable equilibrium. The parameters of this noise are chosen to be close to experimentally relevant parameter ranges and noise structures, while remaining sufficiently close to the ground state. We then simulate the evolution of the system through the eGPE (unitary evolution) and implement a linear ramp of the scattering length into the supersolid regime to $a_s^{f}=88\,a_0$, i.e., $a_s(t)/a_0=96-8\,t/\tau_Q$, with the ramp time set by $\tau_Q$, which takes values between $1\,$ms and $770\,$ms. A schematic of the quench protocol is shown in Fig.~\ref{fig:kzfig} (d). By including stochastic noise in the initial state, and performing 500 independent simulations with identical quench parameters, the dynamics are simulated within a truncated Wigner framework, thereby going beyond a pure mean-field analysis \cite{Blakie2008}.
	
	The variational parameters are fixed at $\ell=1.08\,\mu\text{m}$, $\eta = 4.25$, corresponding to the values in the ground state at $88\,a_0$. The total system size is selected to be approximately $L=344\,\mu$m such that the ground state achievable by perfect adiabaticity would consist of exactly $N_{\text{uc}}=128$ primitive unit cells of the supersolid crystal. See also the Appendices for results with zero temperature (quantum) noise only (App.~\ref{app:SMnoise}), and at larger system sizes (App.~\ref{app:SMLong}).
	
	For the fixed values of the variational parameters above, the critical point of the system is located at $a_{\mathrm{c}}\approx 91.05\,a_0$. It should be noted that this is slightly shifted from the ground-state critical point $a_{\mathrm{c}}^{\mathrm{(GS)}}\approx90.5\,a_0$ (see previous section, Fig.\ \ref{fig:kzfig} (b), and Ref.\ \cite{blakie2020supersolidity}). This shift is expected since $\ell$ and $\eta$ vary slightly across the phase diagram and do not remain fixed. We do not expect this effect to have significant impact on our results due to the continuous character of the transition. The shift of the critical point due to fixing the variational parameters is less than the shift set by using the dimensionally reduced theory. In App.~\ref{app:SM3D}, we show some comparisons of our results with full 3D simulations.
	
	\begin{figure}[t]
		\centering
		\includegraphics[width=0.9\columnwidth]{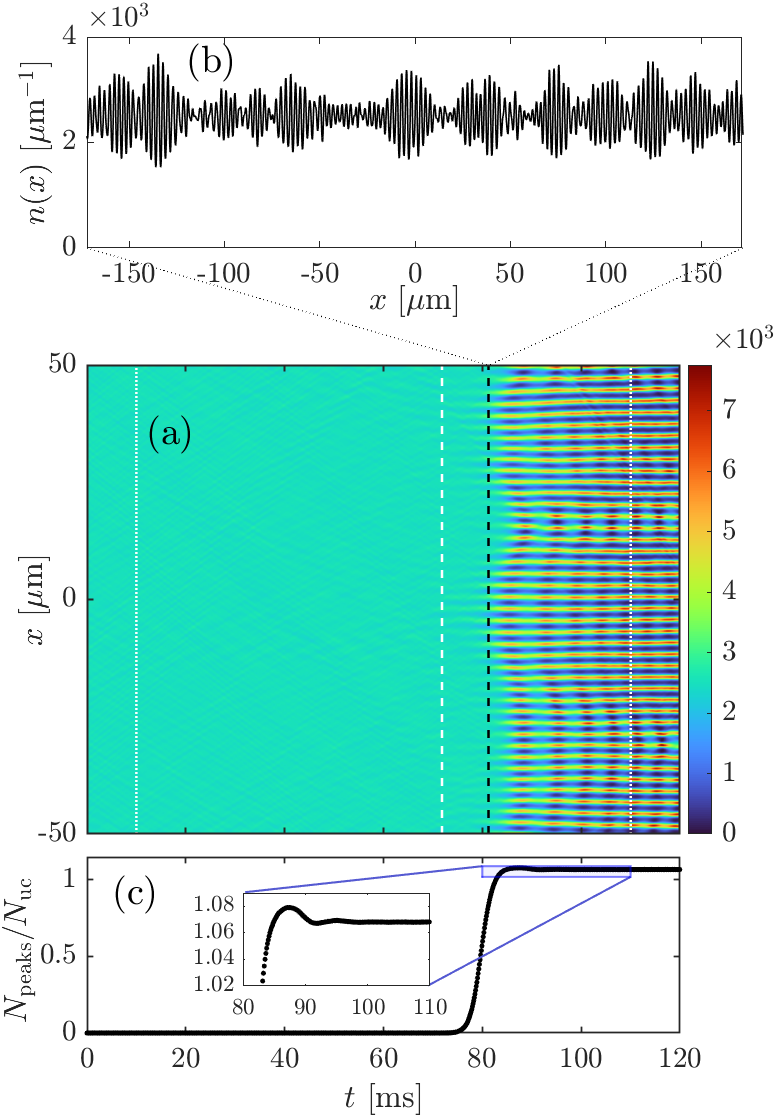}
		\caption{Quench across the superfluid-supersolid transition. Panel (a) shows the density $n(z)$ of a part of the system, for a sample quench, equilibrating for $10\,$ms (leftmost dotted line), then ramping across the critical point (white dashed line) for $100\,$ms until the final value of $88\,a_0$ (right dotted line) is reached. A density cut along the black dashed line is shown in (b) for the full system. Panel (c) shows the number of supersolid peaks per ground-state unit cell as a function of time for the same quench rate as (a)-(b), averaged over 500 independent noise realizations, with an inset showing an overshoot and rebound following the transition.}
		\label{fig:QuenchPic}
	\end{figure}
	
	Fig.~\ref{fig:QuenchPic} shows a sample quench across the superfluid-supersolid transition for $\tau_Q=100\,\mathrm{ms}$. In Panel (a), a portion of the 1D density as a function of time is shown. Density modulations emerge after the critical point is crossed with an apparent delay compared to the expectation for an adiabatic crossing of the transition \cite{Alana2023,Alana2024}. The nonuniformity of the density modulation front suggests that independent regions begin to form domains at different times. This is further emphasized in panel (b), which shows the full 1D density profile with clearly identifiable regions where the modulation has developed only locally. Panel (c) shows the number of density peaks as a function of time (averaged over 500 independent runs) relative to the expected number in the ground state (128 in this case), showing a smooth development of modulations after the transition is crossed. Furthermore, the quench appears to produce an excess of density modulation peaks, first overshooting the long-time average, and then settling to a value above unity within the time scales we consider here. The number of peaks does not settle down to the expected $N_{\text{peaks}}=N_{\text{uc}}$ within at least a hundred milliseconds, indicating that some frustration remains after the quench has been completed, and thus some excess energy remains. Such frustration has also been observed in phase transitions of other superfluids where coarsening dynamics does not occur in low dimensional configurations \cite{Sabbatini2011,Wheeler2025}.

	\section{Scaling Properties}
	\label{sec:scaling}
	
	\subsection{Freeze-out time}
	\label{subsec:freeztime}
	
	According to KZM theory, for a linear quench across a continuous phase transition, the freeze-out time is known to scale as $\hat{\tau}\propto \tau_Q^{\zeta_\text{KZ}}$\;, where $\zeta_\text{KZ}=z\nu/(1+z\nu)$ as in Eq.\ \eqref{eq:kzscaling}. In our system, this indicates that there will exist a quench rate-dependent delay between the crossing of the phase transition and formation of a density modulated superfluid, indicative of a supersolid state. 
	
	\begin{figure}[t]
		\centering
		\includegraphics[width=0.9\columnwidth]{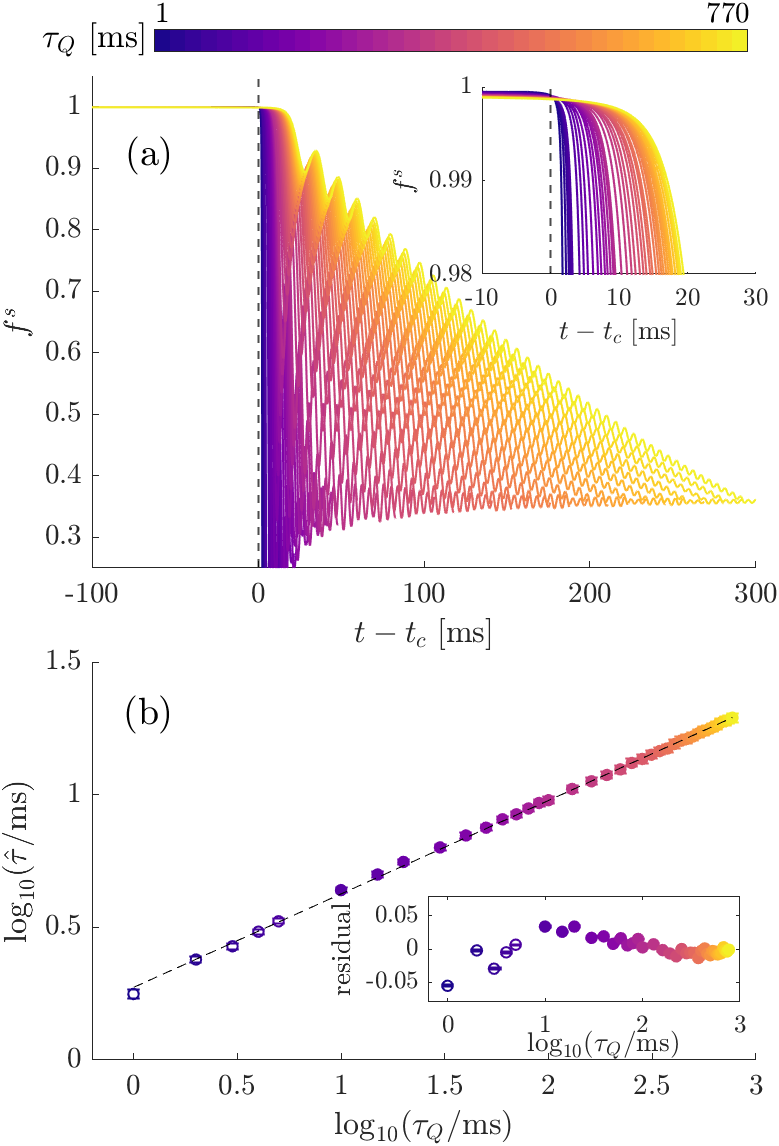}
		\caption{Delay in supersolid formation for various quenches shown via the decrease of $f^s$ from unity. Panel (a) shows the full evolution, with the inset focusing on the initial one-dimensional supersolid formation. Slower quenches lead to a greater delay in the formation of a supersolid following the crossing of the critical point. In (b) the time delay $\hat{\tau}$, at which the superfluid fraction drops below 0.98, is shown as a function of the quench time $\tau_Q$, with the best fit  $\hat{\tau}\propto \tau_Q^{0.352(3)}$ shown as a dashed straight line. Standard errors on the data are shown due to the statistical spread over 500 trials. The inset shows the normalized residuals from the data to the best fit. Open markers represent quenches past $88\,a_0$ (see main text).}
		\label{fig:freezeout}
	\end{figure}
	
	We characterize the emergence of the density modulation through the evolution of the superfluid fraction, estimated via Eq.\ \eqref{eq:Leggett} and ensemble averaged over 500 realizations. In Fig.~\ref{fig:freezeout}, we show the dynamics of $f^s$, which we take to be an order parameter for the superfluid-supersolid transition. In panel (a), we plot $f^s$ across the quench, shifted by the time at which the transition is crossed, $a_s(t_{\mathrm{c}})=a_{\mathrm{c}}$, so that all quenches cross the transition at $t-t_{\mathrm{c}}=0$. After a delay, density modulation develops in the system leading to a depression of the superfluid fraction. This reduction is non-monotonic, and there are oscillations in $f^s$ as the system exhibits a rebounding of the spatial density modulations. This oscillation suggests that an amplitude mode is excited, an example of which is the gapped Higgs excitation that emerges from the roton instability; see Sec.\ \ref{subsec:scaling} and Ref.\ \cite{Blakie2023} for a more in-depth discussion of the emergence of this Higgs mode. In all quenches, there is a final trend towards the ground state superfluid fraction of approximately $f^s=0.323$, settling slightly above this value with some oscillations remaining, indicating some residual excitations (see also discussion of Fig.~\ref{fig:QuenchPic} (c)).
	
	In order to avoid the influence of the excited amplitude mode and any coarsening effects it may have on the system, in this section we consider only the onset of supersolid formation by taking $\hat{\tau}$ to be defined by the difference between the time at which the critical point is crossed to when $f^s$ first drops below 0.98, see inset. Using this definition of $\hat{\tau}$, in panel (b) we show the scaling as a function of the quench time $\tau_Q$. We see clear power-law scaling in agreement with Eq.\ \eqref{eq:kzscaling} and fit $\hat{\tau}\propto \tau_Q^{0.352(3)}$. For the data in this and subsequent figures, we also plot the normalized residual, for Fig.~\ref{fig:freezeout} defined by $(\hat{\tau}-\text{fit}[\hat{\tau}])/\text{fit}[\hat{\tau}]$, where $\text{fit}[\hat{\tau}](\tau_{Q})=A\cdot\tau_{Q}^{\zeta_\text{KZ}}$ is the best fit of $\hat{\tau}(\tau_{Q})$ in terms of a pure power law, parametrized by a real constant $A$ and an exponent $\zeta_\text{KZ}$. In App.~\ref{app:SMUniversal}, we also demonstrate that the extracted $\zeta_\text{KZ}$ exponent can be used to rescale $f^s(t)$ at early times onto a single universal scaling function.

    We select a broad range of quench times to cover what might be considered `reasonable' within an experimental setting, however it should be noted that for inverse rates faster than roughly $\tau_Q\approx 10\,$ms, the quench must be extended past the selected final $a_s= 88\,a_0$, and eventually into the insulating droplet phase (for all these points we choose a final $a_s=60\,a_0$). This is because for very fast quenches, the finite width of the supersolid region would require stopping the ramp during the freeze-out time, resulting in an effectively nonlinear ramp. The quench protocol shown in the schematic of Fig.~\ref{fig:kzfig} (d) becomes modified to follow the dashed line instead -- starting and stopping at higher and lower values of the scattering length, respectively. Our measures of KZM should not depend on this fact, since we only concern ourselves with initial supersolid formation following the crossing of the critical point. However it is important to note that from a practical point of view, there is, generically, an upper limit to how fast one can quench into a supersolid. Throughout the present work, we mark the data obtained by quenches that are taken past the supersolid phase boundary, by open symbols.
	
	\subsection{Correlation length}
	\label{subsec:corlength}
\begin{figure}
		\centering
		\includegraphics[width=0.9\columnwidth]{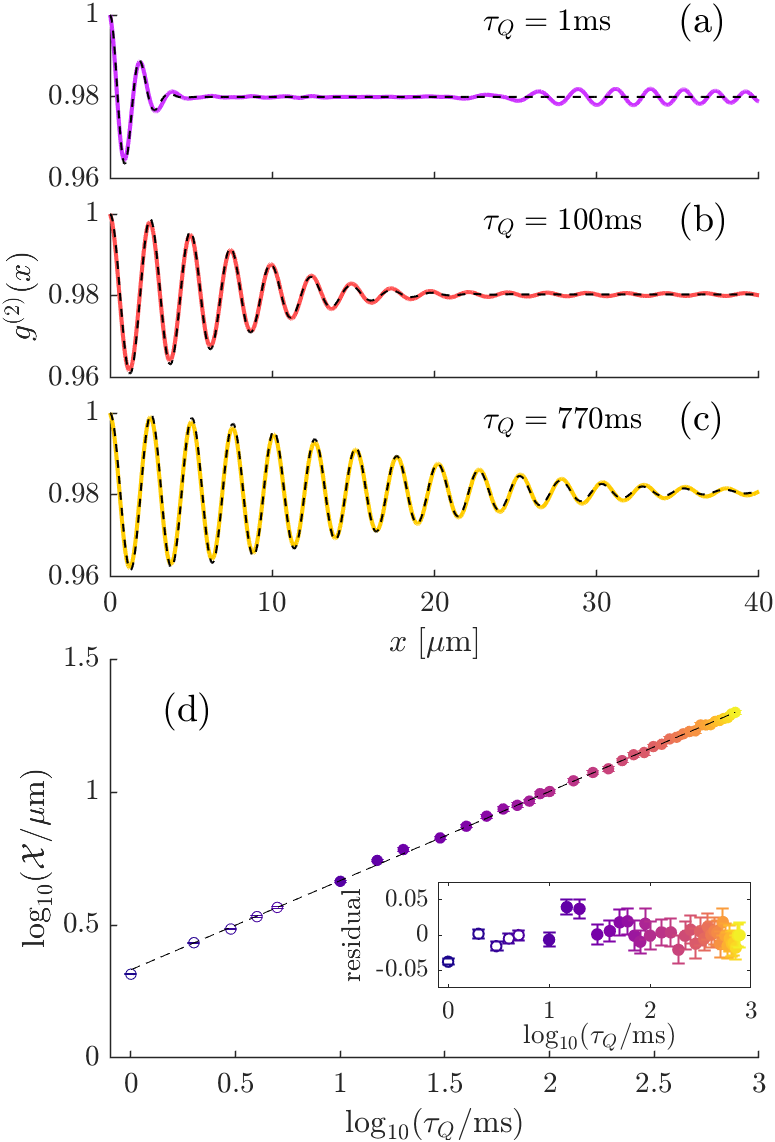}
		\caption{Scaling of the correlation length. Panels (a)-(c) show the correlation function $g^{(2)}(x)$ (solid) for $\tau_Q=\{1,100,770\}$ ms, respectively. In each panel, $g_{\text{fit}}^{(2)}(x)$ is plotted as a black dashed line. 
		Panel (d) shows the width $\mathcal{X}$ for the fit $g_{\text{fit}}^{(2)}(x)$ for all quench durations. The best fit $ \mathcal{X}\sim \tau_Q^{0.335(3)}$ is shown as a dashed straight line. Colors and marker filling match Fig.~\ref{fig:freezeout}. The inset shows the normalized residuals $(\mathcal{X}-\text{fit}[\mathcal{X}])/\text{fit}[\mathcal{X}]$  of the best power-law fit $\text{fit}[\mathcal{X}](\tau_{Q})=B\cdot\tau_{Q}^{\nu_\text{KZ}}$ of $\mathcal{X}$.
        }
		\label{fig:xi_GC}
	\end{figure}

	During a linear quench, the correlation length remains frozen at a value $\hat{\xi}$ when entering the freeze-out regime, which, according to KZM theory, scales as $\hat{\xi}\propto \tau_Q^{\nu_\text{KZ}}$\;, with $\nu_\text{KZ}=\nu/(1+z\nu)$
	cf.~Eq.~\eqref{eq:kzscaling}. Typically, the one-body density matrix $g^{(1)}(x,x')=\langle\psi^*(x)\psi(x')\rangle$ is used to characterize the superfluid nature of condensates \cite{Pitaevskii16,Boninsegni2012a} (i.e.~off-diagonal long-range order) and can be used to define a correlation length associated with the typical range of phase coherence.
    Since the transition to a supersolid is signaled by a growth of periodic diagonal long-range order via density modulations in a system that remains largely phase coherent on both sides of the transition, it can be characterized by the density-density correlation function, viz.,
	\begin{equation}
		g^{(2)}(x-x')=\frac{1}{\bar{n}^2}\langle\psi^*(x')\psi^*(x)\psi(x)\psi(x')\rangle\;, \label{eq:g2}
	\end{equation}
	where we take $\langle...\rangle$ to indicate an ensemble average, in this case corresponding to a statistical numerical average over 500 independent realizations of the initial noise. Translation invariance allows us to interpret the ensemble average also as a spatial average and so this function is evaluated directly in Fourier space via $g^{(2)}(x)=\mathcal{F}^{-1}\left[n(-k)n(k)\right](x)$. In Fig.~\ref{fig:xi_GC}(a)-(c), the solid coloured lines show this function for three sample quench rates at the onset of supersolid formation (i.e., when the superfluid fraction is crossing $f^s(t)=0.98$).
	
	Due to the periodic crystalline structure, $g^{(2)}$ is highly oscillatory while also decaying towards the background superfluid level at that instant. The decay of the initial oscillations indicates that there exists a range over which the supersolid has formed a regular density structure. We find that the numerically extracted $g^{(2)}$ is empirically well-approximated by the function,
	\begin{equation}
		\text{fit}[g^{(2)}](x)= \mathcal{A}+(1-\mathcal{A})\cos\left(\mathcal{K}x\right)\mathrm{e}^{-x^2/\mathcal{X}^2}\;,\label{eq:g2GC}
	\end{equation}
	where $\{\mathcal{A},\mathcal{K},\mathcal{X}\}$ are fitting parameters. In general, we expect $\mathcal{A}$ to match closely with the average background superfluid value $f^{s}(t)$, while the wavevector $\mathcal{K}$ of the modulation in $g^{(2)}$ captures the wavevector of the state's density modulation and should be close to the roton wavenumber $k_{\text{rot}}$ that triggers the instability. The decay of supersolid correlations is set by the width $\mathcal{X}$, which we expect to scale as the correlation length. 
	
	In Figs.~\ref{fig:xi_GC} (a)-(c), the best fit of the function $g^{(2)}(x)$ is shown as a black dashed line in each panel. Panel (d) shows the fit parameter $\mathcal{X}$ over all quenches: we are able to observe a clear power-law scaling in agreement with Eq.\ \eqref{eq:kzscaling}, and plot $\log_{10}(\mathcal{X}/\mu\text{m})$ together with the corresponding KZM power-law fit, $\mathcal{X}\sim\hat{\xi}\sim \tau_Q^{0.335(3)}$.
	Similar scaling can be extracted by simply considering the width of a root-mean squared envelope of $g^{(2)}(x)$ as the oscillations decay (not shown here). The decaying envelope indicates that regular periodic order is not maintained globally across the crystal, but rather that the system forms locally regular domains that are globally incommensurate. There also appears to be a small revival in supersolid coherence in panel (a) that is not captured by our fitting function. We expect that revivals of this kind become less relevant with larger statistical ensembles. As in the case of the superfluid fraction, we are able to use the extracted $\nu_{\mathrm{KZ}}$ in order to rescale $g^{(2)}(x)$ at different quench rates onto a universal scaling function, shown in App.~\ref{app:SMUniversal}. In this case, it is only the envelope of the oscillations that scales universally, since the typical supersolid oscillation period is set by the roton.
	
	The Gaussian spatial decay of the supersolid order is also notable, since generically order parameters will decay exponentially in the non-critical regime. Exceptions exist, for example in free-fermion spin chains, where interdefect correlators follow Gaussian decay \cite{Roychowdhury2021,Nowak2021,Kou2022,Kou2023} following a quench. Non-conservative fields undergoing phase-ordering (e.g., in the 1D XY model) at late times \cite{Bray1994}, beyond the Kibble-Zurek regime, can also yield transient Gaussian correlators.
	
	\subsection{Crystal phase and defect densities}
	\label{subsec:defects}
	
	Another way to define supersolid correlation is to identify a spatially-varying crystal phase along the full system and evaluate the phase coherence over space. This definition is instructive and allows us to directly identify the defects in the crystal structure through phase jumps. Below we describe how to define the crystal phase, check the scaling of the correlation length through that of the crystal phase coherence, and study the scaling of the number of defects.
	
\begin{figure}
		\centering
		\includegraphics[width=0.9\columnwidth]{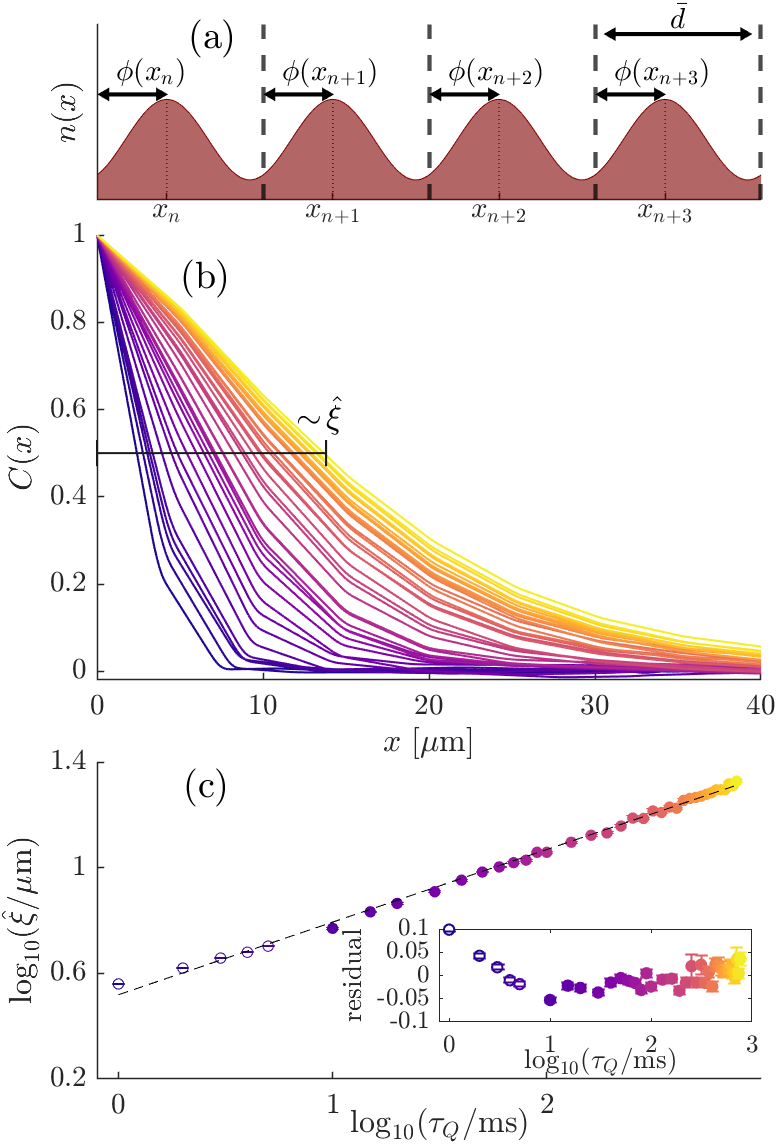}
		\caption{Crystal phase and correlation function $C(x)$. A schematic for the crystal phase is shown in (a), where a modulated supersolid density structure with (in general unevenly spaced) peaks at $x_{j}$, marked with dotted lines. The normalized distance from those peaks to some reference set of points equally spaced by $\bar{d}$, shown as vertical dashed lines, gives the crystal phase $\phi(x_j)$ at that peak. 
		Panel (b) shows the correlation function \eqref{eq:crystalcorr} for the same quenches as used for Fig.~\ref{fig:freezeout}. The half width at half maximum is depicted in panel (c), together with the power-law fit $\hat{\xi}\propto \tau_Q^{0.275(3)}$, with the normalized residuals computed in analogy to those in Fig.~\ref{fig:freezeout} and shown in the inset. Colors and marker filling for (b)-(c) match those in Fig.~\ref{fig:freezeout}.}
		\label{fig:xi_fwhm}
	\end{figure}

	The local phase $\phi$ of the crystal at a point $x=x_j$ can be defined in the following way,
	\begin{equation}
		\phi(x_j)=\frac{2\pi}{\bar{d}}\left(x_j-j\bar{d}\right)\,,
		\label{eq:CrystalPhase}
	\end{equation}
	where $\bar{d}={N_{\text{peaks}}^{-1}}\sum_j'(x_{j+1}-x_j)$ is the average peak-to-peak distance, and the prime indicates that the sum is taken over the density peaks, numbered from $j=1$ to $N_{\text{peaks}}$. This phase is schematically depicted in Fig.~\ref{fig:xi_fwhm} (a), defining $\phi(x_j)$ as the normalized distance of the individual peaks from their respective adjacent minimum in some reference lattice with lattice constant $\bar{d}$ (vertical dashed lines). In analogy with a perfectly coherent uniform superfluid, a globally coherent crystal would therefore have $\phi(x_j)=\ $constant. Numerically, this phase is the phase factor of a Fourier transform of each unit cell of size $\bar{d}$.
	
	The supersolid coherence can also be quantified via its corresponding crystal phasor $\mathrm{e}^{\mathrm{i}\phi(x)}$ and its two-point correlation function,
	\begin{equation}
		C(x-x')=\langle\mathrm{e}^{-\mathrm{i}\phi(x)}\mathrm{e}^{\mathrm{i}\phi(x')}\rangle\;.
		\label{eq:crystalcorr}
	\end{equation}
	Unlike for $g^{(2)}$, the phase $\phi(x)$ is, in each run, only evaluated at the (randomly located) positions $x_j$ of the density maxima, which results in a much coarser structure. Fig.~\ref{fig:xi_fwhm} (b) shows the crystal correlation function \eqref{eq:crystalcorr} following the quench, at the moment $t$ when the superfluid fraction drops below $f^s(t)=0.98$, i.e., at $t=\hat{\tau}$. The width at half maximum of $C(x)$ is proportional to the correlation length $\hat{\xi}$ at the freeze-out time and is then plotted as a function of quench time in panel (c). The extracted correlation length shows power-law scaling, $\hat{\xi}\propto \tau_Q^{0.275(3)}$, a somewhat smaller estimate than the scaling of $g^{(2)}$. In App. \ref{app:SMg2Cz}, we discuss the close relationship between $C(x)$ and $g^{(2)}(x)$.
    
    In computing $C(x)$, it is the averaging over hundreds of trials that ultimately smooths out the data. From the resulting averages one obtains a smooth function that nevertheless develops kinks where its slope changes suddenly due to the higher likelihood of finding a lattice site below that distance. Importantly, at faster quenches, the separation of scales begins to break down: the width of $C(x)$ becomes on the order of the density modulation periodicity $L_{\text{uc}}\approx 2.6\,\mu\mathrm{m}$. Without the separation of scales, the notion of a KZM domain and a supersolid defect become poorly defined \footnote{We therefore suggest that care must be taken when using a function like $C(x)$ to determine KZM scaling results, especially at quench rates where $\xi$ begins to approach $L_{\text{uc}}$, and perhaps greater statistics are needed for proper convergence of this measure.}. 
	
	The KZM is often associated with the formation of topological defects \cite{Uhlmann2010a,Uhlmann2010b}, which in the case of uniform superfluids typically take the form of vortices or solitons \cite{Sadler2006,Uhlmann2007,Zurek2009,delcampo2011,delcampo2013,bland2020persistent,Liu2020}, presenting themselves as jumps in the phase, ultimately separating regions of different gauge symmetry. KZM-induced supersolid defects in the one-dimensional system can be understood instead as sudden jumps in the \textit{crystal} phase $\phi$, noted $\Delta\phi$. Based on the fundamental predictions of KZM theory quoted above, the defect density can be estimated to scale as,
	\begin{equation}
		n_{d}\sim \frac{\xi^d}{\xi^D}\sim \left(\frac{1}{\tau_Q}\right)^{(D-d)\frac{\nu}{1+z\nu}}\equiv \tau_Q^{(d-D)\nu_\text{KZ}}\;, \label{eq:defectdensity}
	\end{equation}
	where $d$ and $D$ are the dimensionality of the defects and system, respectively. In the case of sudden kinks in the crystal phase alone, we therefore expect our results to correspond to $d=0$ and $D=1$, indicating that the defect density scaling should match that of $\hat{\xi}$.
	
	\begin{figure}
		\centering
		\includegraphics[width=0.9\columnwidth]{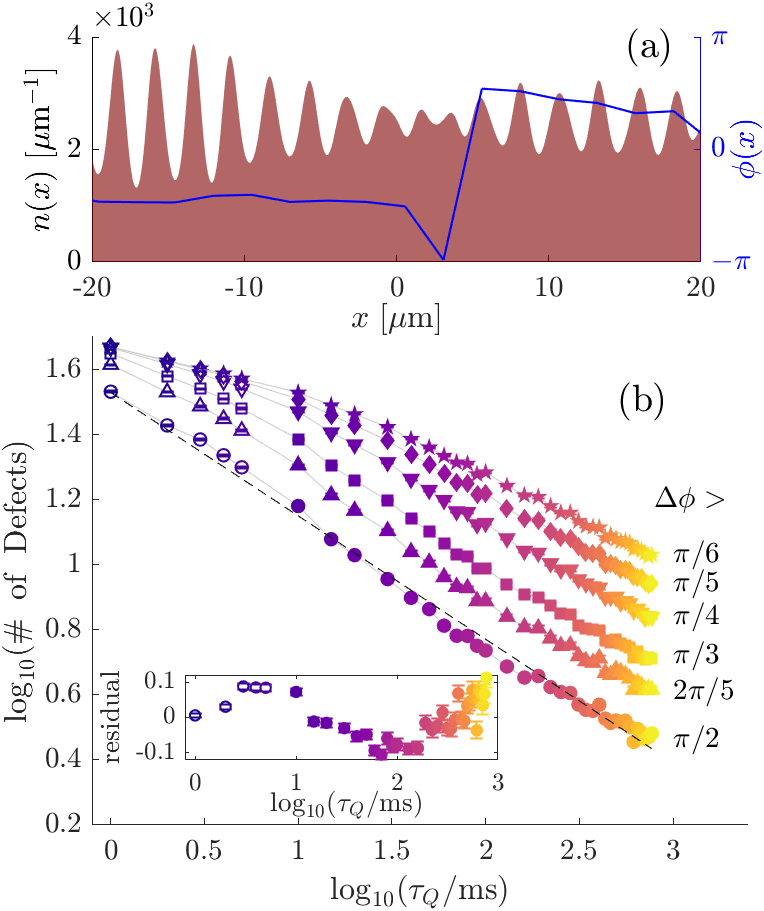}
		\caption{Defect density scaling. Panel (a) illustrates a crystal defect from one simulation run in the density $n(x)$ (brown shaded area) and the crystal phase $\phi(x)$. Defects are identified as local jumps in the crystal phase $\phi(x)$. Panel (b) reports on the number of crystal defects observed at the supersolid formation time for different minimum thresholds of phase jumps, $\Delta \phi> \{\pi/6,\pi/5,\pi/4,\pi/3,2\pi/5,\pi/2\}$, marked as \{circles, triangles, squares, inverted triangles, diamonds, stars\}, respectively. Light gray lines along data sets with the same defect size are shown to guide the eye.  The best fit line for $n_d\propto \tau_Q^{-\nu_\text{KZ}}$ is shown for $\Delta \phi >\pi/2$ only (dashed). Normalized residuals determined in analogy to those in Fig.~\ref{fig:freezeout} are shown in the inset. Colors and marker filling for (b) match those in Fig.~\ref{fig:freezeout}.}
		\label{fig:defects}
	\end{figure}
    
	In Fig.~\ref{fig:defects}, we plot the number of defects ($\propto n_d$) as a function of quench rate. Due to the continuous nature of $\phi$, there remains some ambiguity as to the required size of crystal phase jump to correspond to a defect. We therefore show the scaling for a range of phase jumps corresponding to $\Delta \phi> \{\pi/6,\pi/5,\pi/4,\pi/3,2\pi/5,\pi/2\}$. In all cases, a clear scaling of the number of defects with the quench rate is observed. At faster quench rates, the data appear to curve and saturate towards a behavior that indicates that the defect density begins to be independent of the quench rate. In our case, we again attribute this to the breakdown of scale separation in our system: typical KZM domain sizes begin to approach the crystal wavelength, below which any notion of a supersolid defect becomes poorly defined. This saturation behavior has also been seen in other systems \cite{Donadello2016,Ko2019,Goo2022} and has been described in terms of a universal breakdown \cite{Zeng2023} of the KZM at fast quench rates.
	
	In order to select an appropriate definition of a crystal defect, we consider their number distribution statistics. In what has become known as beyond-KZM physics \cite{delCampo2021,Balducci2023}, universality in defect formation is not restricted to their mean number, but the distribution itself. Since defect formation due to the KZM corresponds to 
	statistically independent events (random successes of Bernoulli trials at different points in space), the discrete probability distribution is binomial \cite{delCampo2018,GomezRuiz2020}. In the continuum limit, where the defects are sufficiently dilute to remain uncorrelated, we therefore expect the probability of counting $\mathcal{N}$ defects to obey a Poisson distribution \cite{delCampo2021,Cui2020,Bando2020},
	\begin{equation}
		P(\mathcal{N})=\frac{1}{\mathcal{N}!}{\lambda}^\mathcal{N}\mathrm{e}^{-{\lambda}}\,,
	\end{equation}
	where expected value $\lambda=\langle\mathcal{N}\rangle$ is then set by the quench rate according to the standard KZM scaling.
	
	\begin{figure}[h]
		\centering
		\includegraphics[width=\linewidth]{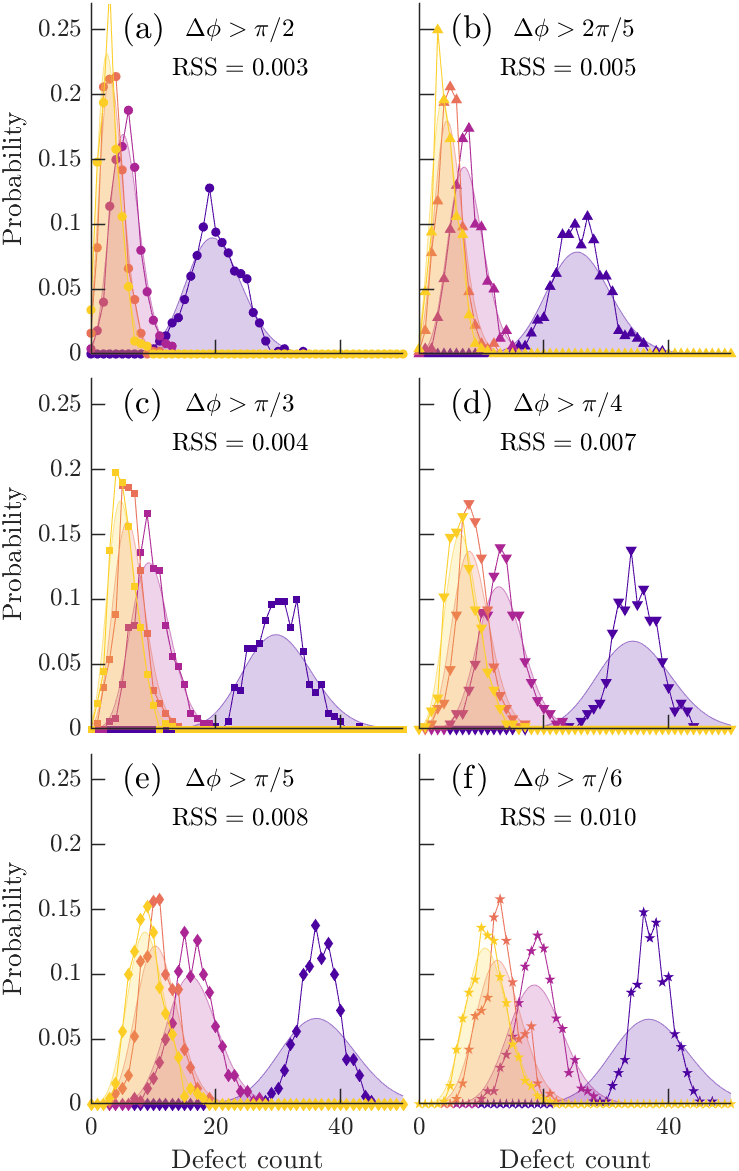}
		\caption{Defect counting statistics. The probability of occurrence of a certain number of phase jumps within a fixed system size $L$ is plotted (filled symbols) for different $\tau_Q \in \{10,100,404,709\}$\,ms, and different $\Delta \phi$ (see panel captions. Symbols match those in Fig.~\ref{fig:defects}). The Poisson distribution corresponding to the maximum likelihood estimate for $\lambda$ for each dataset is shown as a shaded area. Lines serve as a guide to the eye. The residual sum of squares (RSS) is quoted in each panel. Colors match those in Fig.~\ref{fig:freezeout}}
		\label{fig:Poisson}
	\end{figure}
	
	For a selection of defect sizes and quench rates, we plot, in Fig.~\ref{fig:Poisson}, the defect count probabilities within our total system following a quench, and compare them to the best-fit Poisson distribution curves $P(\mathcal{N})$. The difference of the curves is quantified through the average sum of squared residuals (RSS). A stricter definition of a defect (i.e., larger $\Delta\phi$) leads to better agreement (smaller RSS) with Poisson distributions across different $\tau_Q$, while smaller phase jumps yield narrow peaks that disagree with the Poissonian hypothesis, especially at fast quench rates. Deviations from Poisson statistics imply that the phase jumps being counted no longer result from independent random events and thus are no longer `defects' in the KZM sense, but are somehow correlated. We suspect that small changes in the crystal regularity may not only arise due to KZM scaling, but as excitations in the crystal due to propagation of, e.g., sound modes (phonons) within an otherwise coherent domain. 
	
	The analysis of Fig.~\ref{fig:Poisson} indicates that a strict definition of a defect, using a large minimal phase jump $\Delta \phi>\pi/2$ is required for identifying KZM defects. Returning to Fig.~\ref{fig:defects} we fit the $\tau_Q$-scaling of the numbers of defects with $\Delta \phi>\pi/2$ with a power law and extract $\nu_\text{KZ}=\frac{\nu}{1+z\nu}\approx 0.37(2)$. This scaling measure is independent from the $g^{(2)}$ scaling of the correlation length in Sec.\ \ref{subsec:corlength}, and again appears to give a scaling result relatively close to 1/3, albeit it is slightly outside the standard error bounds. We would like to also remark here that considering defects with large phase jumps may introduce a larger statistical spread: With the system sizes we simulate here, slow quench rates may only produce a handful of defects ($\lesssim5$), and so the relative point spread error due to a single defect can be large. In the following section, we shall further examine the compatibility of these results with respect to critical scaling exponents as extracted from Bogoliubov theory.
	
	We note that, in the above analysis, we did not consider the influence of extended ($d=1$) excitations playing a role in supersolid formation. It is known that, for the temperature-quench-induced BEC transition \cite{Donadello2016,Rabga2023}, the interplay between defects of different dimensions (e.g.~vortices vs.~solitons) may hybridize the scaling result of Eq.\ \eqref{eq:defectdensity}. Similarly, any incidence of nonlinear excitations induced by the transition or by the initial thermal fluctuations would therefore, in principle, also contribute to the flattening of the curves in Fig.~\ref{fig:defects}.
	
	\subsection{Scaling exponents}    
	\label{subsec:scaling}
	
	Using the numerical results for $\zeta_\text{KZ}$ and $\nu_\text{KZ}$ from subsections \ref{subsec:freeztime} and \ref{subsec:corlength}, respectively, we extract the scaling exponents,
	\begin{align}
		\nu=&\; 0.57(1)\,,\\
		z=&\;1.05(2)\,,
	\end{align}
	for the superfluid-to-supersolid transition.

\begin{figure}[t]
		\centering
		\includegraphics[width=0.95\columnwidth]{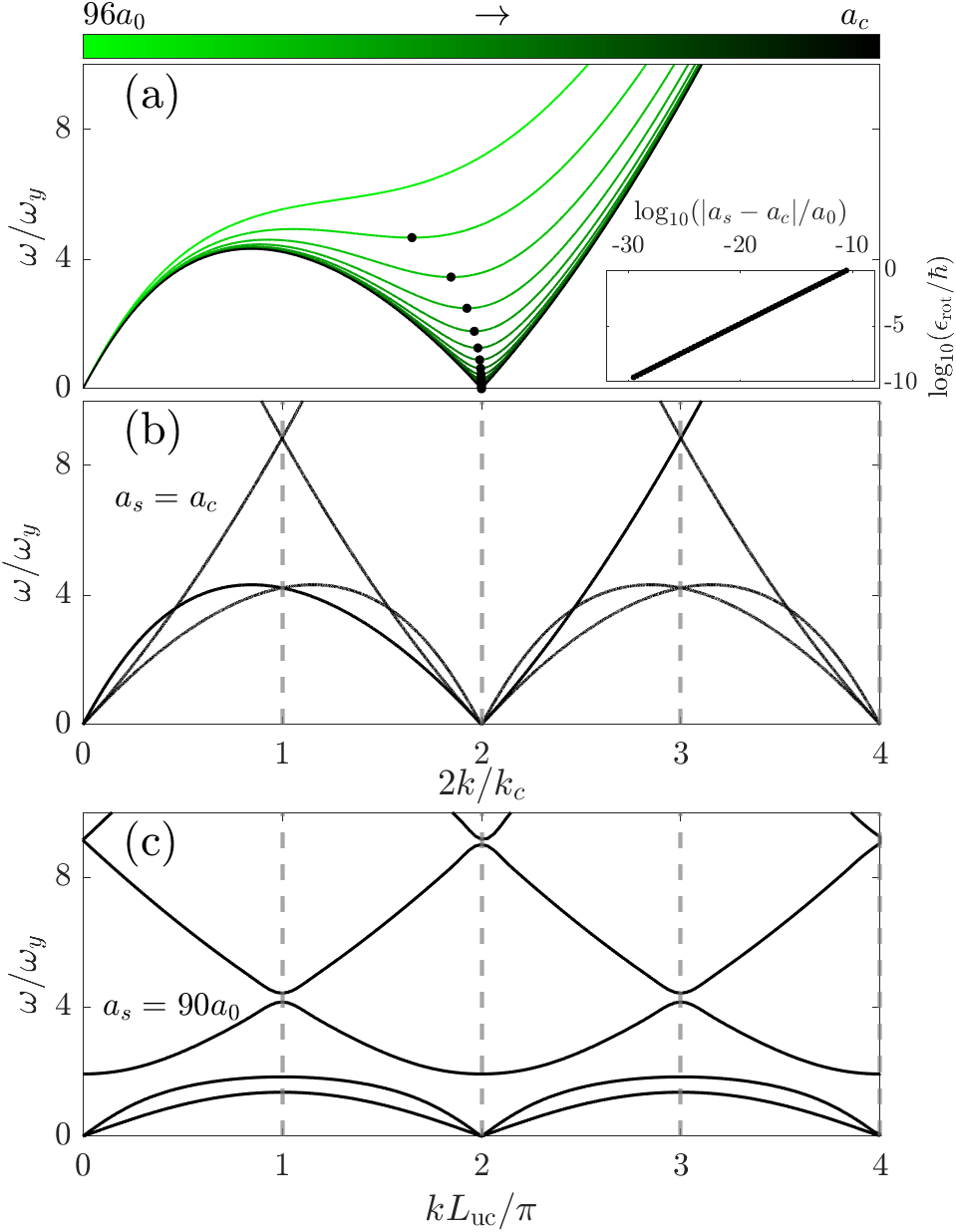}	\caption{Panel (a) show the dispersion relation and the closing of the roton gap from Bogoliubov theory (Eq.~\eqref{eq:Bog}) as $a_s$ is tuned through the superfluid regime. The roton minimum is marked by black circles. The scaling of the roton gap as $a_{s}$ approaches $a_{\mathrm{c}}\approx 91.05\,a_0$ from above is shown in the inset and satisfies $\epsilon_{\mathrm{rot}}\sim \sqrt{a_s-a_{\mathrm{c}}}$. Panel (b) shows the mapping of the spectrum at the critical point [black curve in (a)] into a periodic Brillouin-zone structure with periodicity $k_{\text{rot}}\equiv k_{\mathrm{c}}$. Panel (c) shows the dispersion in the supersolid regime with a periodic Brillouin-zone structure whose periodicity is $2\pi/ L_{\mathrm{uc}}$. BZ edges are marked as vertical grey dashed lines. The excitation frequencies $\omega$ are shown in units of the transverse trapping frequency $\omega_y=\omega_z=2\pi \times 150\,$Hz. 
		}
		\label{fig:RotonSoft}
	\end{figure}

    We may compare these dynamical KZM results to equilibrium estimates within the mean-field approximation. The characteristic time scale of a system near a quantum critical point is determined by the inverse of the energy gap, which closes as $\Delta\sim |\lambda-\lambda_{\mathrm{c}}|^{z\nu}$ near the critical value $\lambda_{\mathrm{c}}$ of some tuning parameter $\lambda$ \cite{sachdev2011QPTbook}. Meanwhile, in the theory of dynamical critical phenomena \cite{Hohenberg1977} the low-energy vs.~low-momentum scaling of the dispersion defines the dynamical critical exponent as $\omega\propto k^z$. In the uniform superfluid, the dispersion relation can be calculated analytically within Bogoliubov theory based on the eGPE \cite{Blakie2020variational}, 
    \begin{equation}
        \hbar\omega=\sqrt{\frac{\hbar^2k^2}{2m}\left(\frac{\hbar^2k^2}{2m}+\frac{6\gamma_\mathrm{QF}n^{3/2}}{5 \pi ^{3/2} \ell^3}+\frac{gn}{ \pi  \ell ^2}+2n\tilde{U}^{\mathrm{1D}}(k)\right)}\label{eq:Bog}\;,
    \end{equation}
    while in the supersolid phase only numerical solutions are possible. At zero temperature and for $a_{s}>a_{\mathrm{c}}\approx 91.05\,a_0$, there exists a single gapless low-energy sound branch (i.e.~Goldstone mode) associated with the phase rigidity of the system. At higher $k$, the attractive contribution of the DDI results in a rotonic minimum at momentum $k=k_{\text{rot}}$ and of energy  $\epsilon_{\mathrm{rot}}$, and it is the softening and instability of the roton mode that leads to supersolid formation, see e.g.~\cite{blakie2020supersolidity}. Fig.~\ref{fig:RotonSoft} (a) shows the dispersion relation in the uniform regime (green curves), with a roton minimum developing as $a_s$ is reduced. In the regime of the phase diagram where the transition is continuous, the onset of supersolidity occurs as the roton minimum touches zero, $\epsilon_{\mathrm{rot}}=0$ (black curve, $a_s=a_{\mathrm{c}}$), and proceeds to go unstable. 
	
	The emergence of periodic density modulation in real space also leads to a periodic dispersion in momentum space and thus to a band structure. Fig.~\ref{fig:RotonSoft} (b) shows the band structure precisely at the critical point, with the Brillouin zone (BZ) edges set at multiples of the roton momentum at criticality, $k_{\text{rot}}\equiv k_{\mathrm{c}}$, see e.g.~\cite{Blakie2023}. This corresponds to the mapping of the dispersion relation at $a_s=a_{\mathrm{c}}$ of panel (a) into the first BZ.  In this mapping, the roton minimum yields a second Goldstone (crystal-sound) mode associated with the breaking of the translational symmetry, and a third branch, which appears gapless at the transition. Precisely at the critical point, the roton wavelength sets the unit cell length via $L_{\text{uc}}=2\pi/k_{\mathrm{c}}$; however, the unit cell size does not remain constant throughout the supersolid regime. Just below the transition, as shown in panel (c) for $a_s=90\,a_0$, the dispersion relation show two gapless Goldstone mode while a gap (re)opens in the first branch at $k=\{0,k_{\mathrm{c}},2k_{\mathrm{c}},...\}$, corresponding to the Higgs mode that drives the amplitude of density modulations. 
	
	Within Bogoliubov theory, we therefore expect the characteristic time to scale inversely to the closing of the rotonic energy gap \cite{sachdev2011QPTbook},
	\begin{equation}
		\tau^{-1}\sim \epsilon_{\mathrm{rot}}\sim |a_s-a_{\mathrm{c}}|^{z\nu}\;.
	\end{equation}
	It has been previously shown \cite{Blakie2020variational}, and is verified here in the inset to Fig.\ \ref{fig:RotonSoft} (a), that the roton gap closes as $\epsilon_{\mathrm{rot}}\sim \sqrt{a_s-a_{\mathrm{c}}}$ so that $z\nu=1/2$. Meanwhile, precisely at the critical point, the dispersion relation linearizes around the roton momentum,
	\begin{equation}
		\epsilon_{\mathrm{rot}}\propto |k-k_{\mathrm{c}}|\;,
	\end{equation}
	and from the BZ mapping, the gapless roton mode emerges periodically about the zone edges, $\epsilon_{\mathrm{rot}}\propto |k-mk_{\mathrm{c}}|,\;m\in\mathds{Z}$. Due to the linearization of the dispersion relation around $mk_{\mathrm{c}}$, for the superfluid to one-dimensional supersolid transition considered here, Bogoliubov theory suggests $z=1$. Interestingly, while the presence of the LHY correction to Eq.~\eqref{eq:1DGPE} and by extension Eq.~\eqref{eq:Bog} does change the location of the critical point slightly, it does not change the exponents extracted using Bogoliubov theory~\cite{Blakie2020variational}. In this sense, the analytic `mean-field' analysis is not changed by the inclusion of a beyond-mean-field correction. A direct numerical analysis without the LHY correction is not possible, since the supersolid state requires stabilization.
	
	The scaling exponents extracted from our numerical simulations, $\nu=0.57(1)$ and $z=1.05(2)$, appear compatible with the mean-field scaling exponent extracted from Bogoliubov theory, $\nu=1/2$, of the correlation length and the quantum-critical value $z=1$ of the dynamical exponent, although these fall slightly outside the statistical error bars of our results. Our numerical simulations go beyond mean-field dynamics, not because of the inclusion of the LHY correction, but because they are performed with a truncated Wigner framework \cite{Steel1998,Sinatra2002,Blakie2008}, which incorporates classical fluctuations in a fully non-perturbative manner. The close agreement between the extracted exponents and mean-field theory may be due to the scaling behavior being relatively independent from beyond-mean-field effects, e.g., as in the case of systems above the upper critical dimension. Due to the relatively low depletion (typically less than 5\% in our simulations at $20\,\text{nK}$ here) we can expect modifications arising from interaction with excitations to be negligible.
	
    A universality class that we had foreseen to apply to our transition is that of the $(D+1)$-dimensional XY model, with $D_\mathrm{lc}=1$ and $D_\mathrm{uc}=3$ being the respective lower and upper critical dimensions. In our case, $D=1$, such that the transition would be of the Berezinskii-Kosterlitz-Thouless (BKT) type \cite{Jelic2011,Dziarmaga2014,Gardas2017,Zuo2021}, for which a non-polynomial correlation length and relaxation time scaling could be expected, which reaches power-law scaling with $\nu\to\infty$, $z=1$ only asymptotically, for exponentially large quench times $\tau_Q$ \cite{Dziarmaga2014}.
    This is in contrast to the clean power-law scaling observed here already for relatively small $\tau_Q$, resulting in the above near-mean-field value of the exponent $\nu$. Furthermore, Blakie \textit{et al.}~\cite{Blakie2023}, who examined the critical behavior at the continuous superfluid-supersolid transition line, using Bogoliubov theory, found that there is a discontinuity in the compressibility, consistent with the transition being second-order (the single exception being at the tip of the phase transition curve, which we do not encounter in any quenches performed in this paper). This discontinuity contrasts with the BKT expectation and supports the hypothesis of a universality class different from (1+1)D XY.  
    Even in $D=3$ spatial dimensions, i.e., at the upper critical dimension of the XY model, the exponents, near the quantum critical point, would receive small anomalous contributions, and $z\simeq (D+1)/2=2$ instead of the exponent $z\simeq1$ we find, which is indeed compatible with dynamical critical scaling in $D=1$ dimensions \cite{Hohenberg1977}. The exact universality class of the transition remains undetermined. 
    We, however, note that our extracted near-mean-field exponents have been previously found to characterize the universality class for the superfluid-to-Mott insulator transition in the Bose-Hubbard model in $D\geq3$ spatial dimensions, when crossing precisely at the Mott lobe tips with integer filling factor, i.e., at the quantum multicritical point~\cite{Fisher1989,Huang2021}, and exponents similar to our KZM results have been previously observed in density-wave order (supersolid) transitions in $D=2$ dimensions~\cite{vanOtterlo1995}. 
 
	\section{Conclusion}
	\label{sec:conclusion}
	
	We have explored the formation of a supersolid in an elongated dipolar quantum gas following a quench of the scattering length, using a reduced 3D model and via the extended Gross-Pitaevskii equation. We extract power-law scaling exponents corresponding to diverging relaxation time and correlation length scales. In particular, the divergence in relaxation time was estimated by identifying the freeze-out time with the delay in supersolid formation during the quench. The supersolid correlation length scaling was extracted by fitting a spatially oscillating function with Gaussian-shape envelope to the density-density correlation function $g^{(2)}(x)$. We found that the results for the extracted critical exponents, $\nu=0.57(1)$ and $z=1.05(2)$, appear to be nearly compatible with the corresponding values expected from mean-field theory. When considering a measure for the crystal phase, $\phi(x)$, we found that there appears to be a systematic breakdown of scaling results and an onset of defect density saturation at very fast quench rates. We posit that this breakdown \cite{Zeng2023} is explained by the fact that typical KZM domain sizes in the supersolid approach the crystal lattice size set by the roton. Nevertheless, over multiple orders of magnitude, defect densities appear to obey power-law scaling.
	
	We consider an elongated system for statistical purposes, however our study is performed with experimental considerations in mind. In particular, we selected parameters corresponding to ${}^{164}$Dy, including scattering lengths, typical trap frequencies, and reasonable temperatures used in dipolar quantum gas experiments \cite{Tanzi2019,Bottcher2019,Chomaz2019}. For a dipolar gas confined in a harmonic trap, the direct applicability of our results will depend on the cloud aspect ratio and quench rates \cite{Liu2020}, since in some regimes the inhomogeneous KZM \cite{delcampo2011,Rabga2023} may lead to a different set of exponents. One candidate for the uniform tubular geometry we have proposed is a ring-shaped trap \cite{Bloch1973,Amico2005,Amico2022}, which has been realized for ${}^{23}$Na \cite{Ryu2007,Ramanathan2011,Wright2013,Eckel2014} and ${}^{87}$Rb \cite{Henderson2009,Sherlock2011,Moulder2012,Ryu2013,Beattie2013,Corman2014,Marti2015}, and has been theoretically shown to support the dipolar supersolid phase \cite{Tengstrand2021,Tengstrand2023}. While supersolid lifetimes remain a challenge for these experiments, particularly due to high densities in the dipolar droplets, the advantage of the measures we have introduced in this paper is that only the onset of supersolidity needs to be considered, thus long lifetimes are not required to verify our results. 
	
	An interesting future direction is to consider melting of the supersolid via a reverse-quench protocol \cite{Mukherjee2008,Divakaran2009,Quan2010,Kou2022}, allowing us to probe the symmetrical of the KZM schematic curves shown in Fig.~\ref{fig:kzfig}. Finally, we have only focused on the transition from the uniform superfluid to one-dimensional droplet chain. If an additional trapping direction is relaxed, there exist transitions to many more kinds of two-dimensional supersolid lattices that are also possible, leaving the door open to further examination of dynamical symmetry breaking in more complex supersolids. 

    \begin{acknowledgements}
        The authors thank M. Ballu, R. Bisset, T. Bland, K. Chandrashekara, J. Gao, C. G\"{o}lzh\"{a}user, P. Heinen, L. Hoenen, G. Lamporesi, A. Oros, E. Poli, G.-X. Su, and N. Rasch for helpful discussions and collaboration on related work.
        They acknowledge support 
        by the Deutsche Forschungsgemeinschaft (DFG, German Research Foundation), through 
        SFB 1225 ISOQUANT (Project-ID 273811115), 
        grant GA677/10-1, 
        and under Germany's Excellence Strategy -- EXC 2181/1 -- 390900948 (the Heidelberg STRUCTURES Excellence Cluster), 
        and by the state of Baden-W{\"u}rttemberg through bwHPC and the DFG through 
        grants 
        INST 35/1134-1 FUGG, INST 35/1503-1 FUGG, INST 35/1597-1 FUGG, and 40/575-1 FUGG
        (MLS-WISO, Helix, and JUSTUS2 clusters). 
        W.K. acknowledges the support of the Natural Sciences and Engineering Research Council of Canada (NSERC), [funding reference number PDF-577924-2023]. L.C. acknowledges support by the European Research Council (ERC) under the European Union’s Horizon Europe research and innovation program under grant number 101040688 (project 2DDip). Views and opinions expressed are however those of the authors only and do not necessarily reflect those of the European Union or the European Research Council. Neither the European Union nor the granting authority can be held responsible for them.  
    \end{acknowledgements}
	\appendix
	\section{Thermal and quantum noise}
	\label{app:SMnoise}
	
	In the variational 1D simulations, thermal and quantum noise is included via,
	\begin{equation}
		\psi(x,0)=\sqrt{\bar{n}}+\sideset{}{'}\sum_{l}\left[A^{+}_l\mathrm{e}^{\mathrm{i}k_l x}+A^{-}_l\mathrm{e}^{-\mathrm{i}k_l x}\right]\,,
		\label{eq:QNoise1D}
	\end{equation}
	where $A^{\pm}_l$ are Gaussian random variables subject to $\langle |A^{+}_l|^2\rangle=\langle |A^{-}_l|^2\rangle=(\mathrm{e}^{\epsilon_l^f/k_\mathrm{B} T}-1)^{-1}+\frac{1}{2}$ for temperature $T$ and free-particle dispersion $\epsilon_l^f=2\hbar^2l^2/(mL^2)$ for $l\in \mathds{N}^+$. The sum $\sum'$ is restricted to eigenstates such that $\epsilon_l^f<2k_\mathrm{B}T$.
	
	In full 3D calculations (see App.\ \ref{app:SM3D}), thermal and quantum noise includes amplitudes in the transverse harmonic oscillator modes,
	\begin{equation}
		\Psi(\textbf{r},0)=\sqrt{\bar{n}}+\sideset{}{'}\sum_{l\gamma\delta}\phi_\gamma(y)\phi_\delta(z)\left[A^{+}_{l\gamma\delta}\mathrm{e}^{\mathrm{i}k_l x}+A^{-}_{l\gamma\delta}\mathrm{e}^{-\mathrm{i}k_l x}\right],
		\label{eq:QNoise3D}
	\end{equation}
	where now $\langle |A^{+}_{l\gamma\delta}|^2\rangle=\langle |A^{-}_{l\gamma\delta}|^2\rangle=(\mathrm{e}^{\epsilon_{l\gamma\delta}/k_\mathrm{B} T}-1)^{-1}+\frac{1}{2}$ for dispersion $\epsilon_{l\gamma\delta}=\hbar\omega_y\left(\gamma-\frac{1}{2}\right)+\hbar\omega_z\left(\delta-\frac{1}{2}\right)+2\hbar^2l^2/(mL^2)$, now with $l,\:\gamma,\:\delta\in \mathds{N}^+$. The sum has again been restricted such that $\epsilon_{l\gamma\delta}<2k_\mathrm{B}T$.
	
	\begin{figure}[h]
		\centering
		\includegraphics[width=0.9\linewidth]{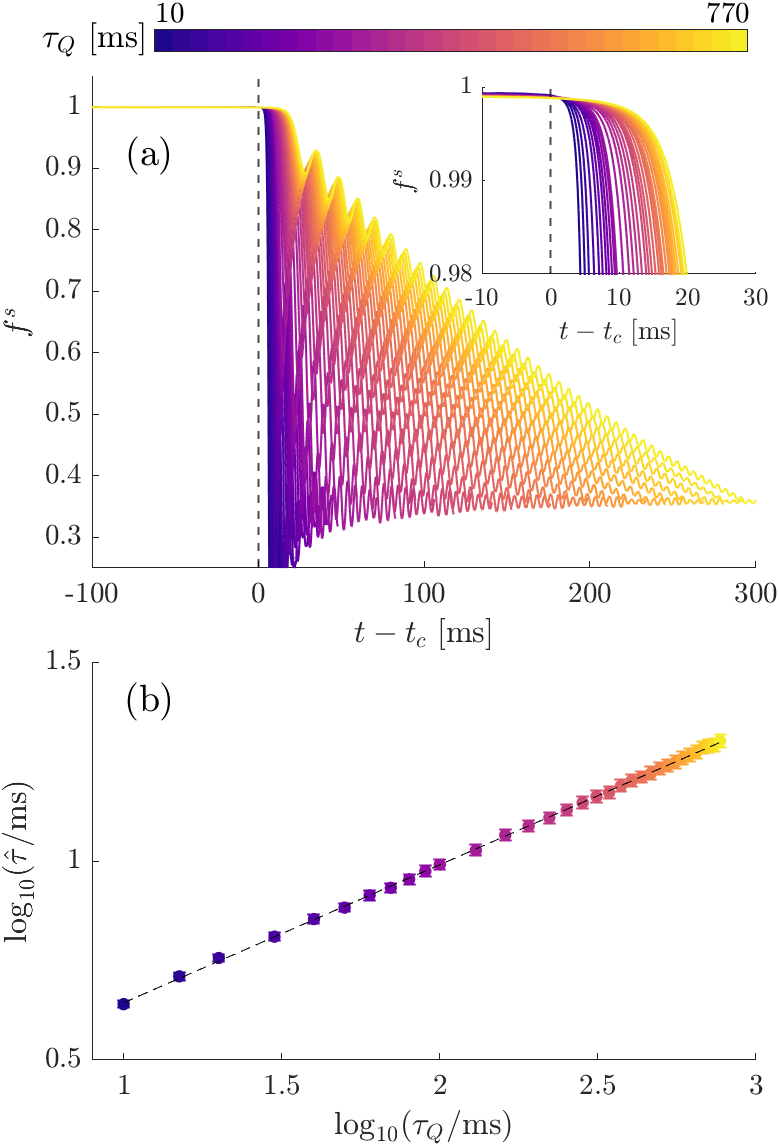}
		\caption{Delay in supersolid formation , similar to Fig.~\ref{fig:freezeout}, with the only difference is each simulation is populated only with quantum noise. As in the main text, (a) shows the full $f^s(t)$ with an inset focusing on the initial moments after crossing the critical point. In (b), we show the time delay, with the best fit $\hat{\tau}_{T=0}\propto \tau_Q^{0.346(2)}$.
		}
		\label{fig:tDelay_QNoise}
	\end{figure}
	
	In the main text, we consider a system populated with stochastic noise drawn from a thermal distribution at 20\,nK. We have assumed that this is close enough to zero temperature such that we are not greatly affecting the location of the transition due to thermal effects. Here, we verify that this assumption holds by considering quenches where the only source of noise is a half-particle per mode. Specifically, we populate modes as in Eq.~\eqref{eq:QNoise1D}, but $A^{\pm}_l$ are now subject to the zero-temperature limit $\langle |A^{+}_l|^2\rangle=\langle |A^{-}_l|^2\rangle=1/2$. Rather than a temperature-dependent cutoff, we now restrict the sum to include states with momenta less than the inverse healing length: $k^{-1}<\xi_{\text{h}}=\hbar/\sqrt{m\mu}$. In this limit, the particle and energy added due to noise has been reduced to less than 1\%.
	
	\begin{figure}[h]
		\centering
		\includegraphics[width=0.9\linewidth]{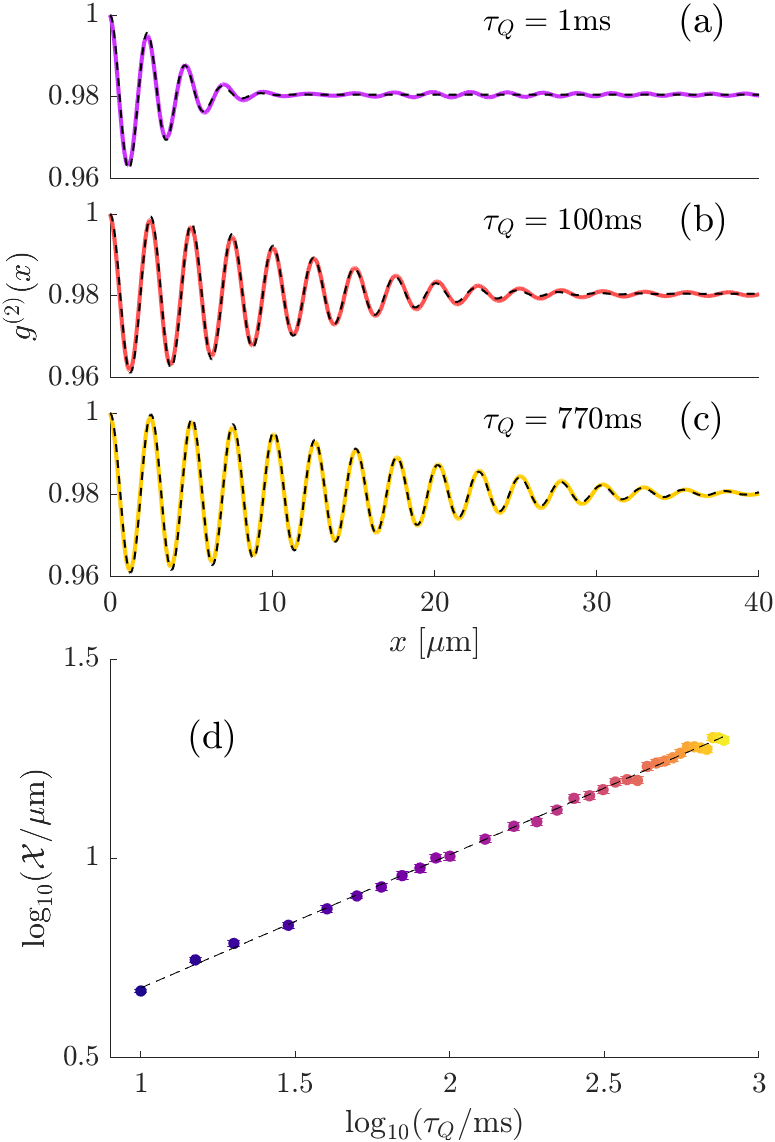}
		\caption{Similar to Fig.~\ref{fig:freezeout} of the main text, now with quenches where the initial state is populated with quantum noise only. Panels (a)-(c) show the density-density correlation function at the indicated times, and (d) shows the envelope scaling as in the main text, with $\mathcal{X}_{T=0}\sim\xi\sim\tau_{Q}^{0.334(5)}$.
		}
		\label{fig:G1G2_QNoise}
	\end{figure}
	
	In Figs.~\ref{fig:tDelay_QNoise} and \ref{fig:G1G2_QNoise}, we present the results for our system populated only with the zero-temperature quantum noise described above, rather than the low-temperature thermal noise of the main text. In these figures, we have selected again the same system size as described in the main text, now averaged over 400 independent trials. The resulting scaling, described in the captions of each figure, seem to indicate that there is very minimal difference due to the choice of noise. This indicates that the $T=20\,\mathrm{nK}$ noise used in the main text is sufficiently small for our parameters to be considered representative of dynamics to be expected in the zero-temperature limit.  
	
	\section{Comparison with 3D simulations}
	\label{app:SM3D}
	
	Here, we compare our results with full 3D simulations, i.e., evolving under Eq.~\eqref{eq:GPE3D}, for select quench rates. Similar to the variational model, we perform quenches from $a_s=96\,a_0$ to $a_s=88\,a_0$, now with a grid size of approximately $\approx 160\,\mu \text{m}$, corresponding to 64 unit cells in the supersolid ground state at $88a_0$. At $\bar{n}=2500\,\mu\text{m}^{-1}$, 3D simulations indicate that the critical point is located at $a_s=a_{\mathrm{c}}^{3\mathrm{D}}=92.27\,a_0$ \cite{Smith2023supersolidity}. Due to the size of the simulation grids and the resulting datasets, the results are averaged over only 5 realizations each of the noise in App. \ref{app:SMnoise}, and for a more limited range of $\tau_Q$, resulting in much larger standard errors in the data.
	
	\begin{figure}[h]
		\centering
		\includegraphics[width=0.9\linewidth]{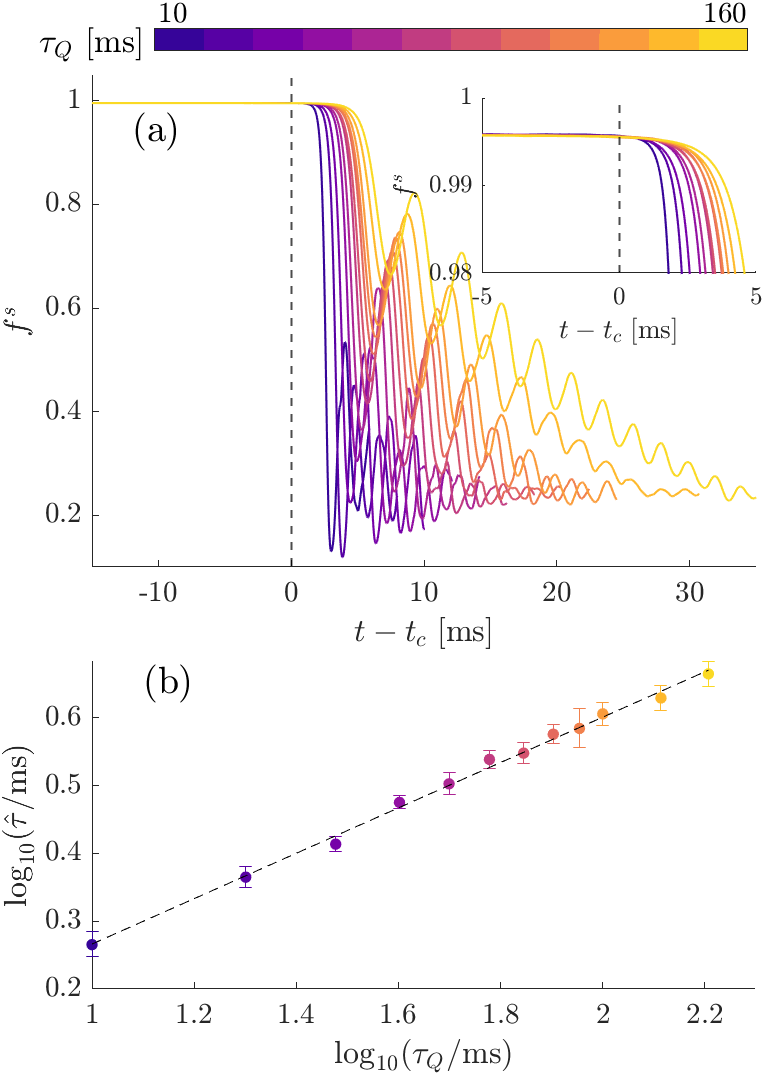}
		\caption{Delay in supersolid formation for various quenches for full 3D simulations. Similar to Fig.~\ref{fig:freezeout} of the main text, panel (a) shows the full $f^s(t)$, with the inset focusing on the initial supersolid formation. Panel (b) shows the time delay at which $f^s$ crosses 0.98, indicating supersolid formation, with the best fit $\hat{\tau}_{3\text{D}}\propto \tau_Q^{0.33(1)}$ shown as a dashed line. Note, the difference in colorbar scale from the other figures due to the limited range of $\tau_Q$.
        }
		\label{fig:tDelay3D}
	\end{figure}
	
	The superfluid fraction $f^s$ as a function of time is presented in Fig.~\ref{fig:tDelay3D} (a). Similar to the main text, we can extract the supersolid delay by the time it takes for $f^s$ to drop below 0.98. The scaling of the delay is plotted in panel (b), along with the corresponding quench data using the variational model. We are able to extract $\hat{\tau}_{3\text{D}}\propto \tau_Q^{0.33(1)}$, which, within the error range quoted, is compatible with our results from the reduced model. 
	
	\begin{figure}[h]
		\centering
		\includegraphics[width=0.9\linewidth]{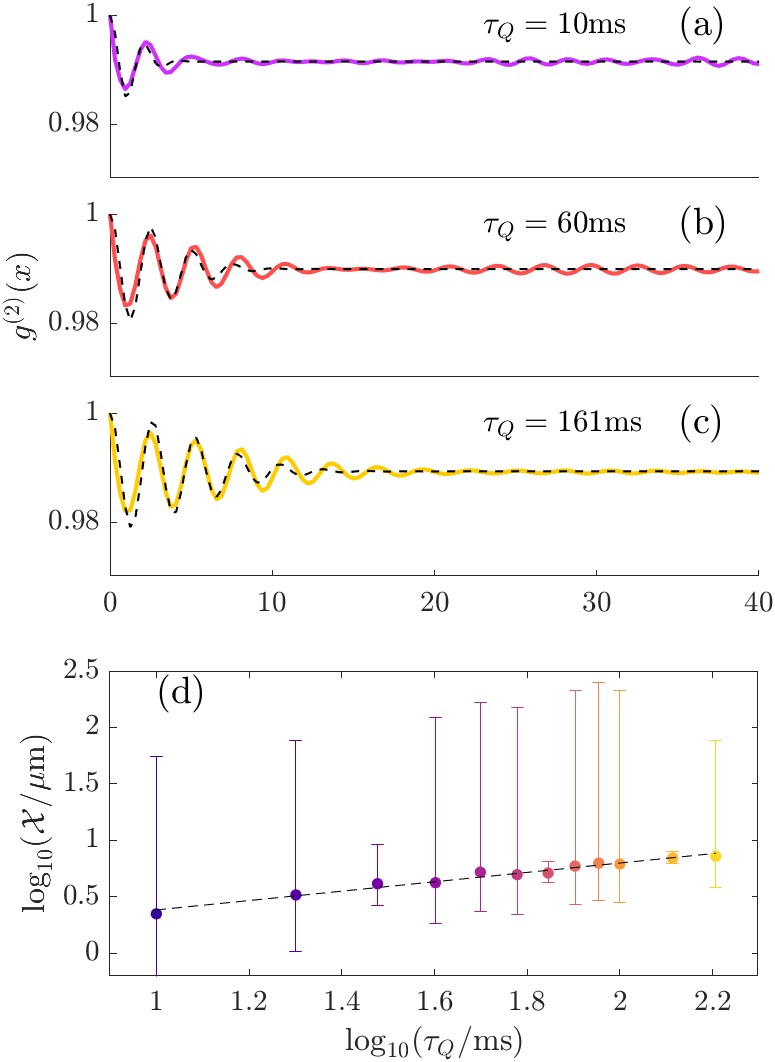}
		\caption{Supersolid correlation length $g^{(2)}(x)$, similar to Fig.\ \ref{fig:xi_GC}, now for 3D simulations, at $\tau_Q=\{10,60,161\}$ in panels (a)-(c), respectively. A giving an approximate scaling of $\mathcal{X}_{3\textbf{D}}\sim\hat{\xi}\sim\tau_Q^{0.40(16)}$.}
		\label{fig:g23D}
	\end{figure}
	
	We can also measure the $g^{(2)}$ correlation function along the $z$-axis for full 3D simulations, as shown in Fig.~\ref{fig:g23D}. Similar to Fig.~\ref{fig:xi_GC} of the main text, we plot in panels (a)-(c) the correlation function $g^{(2)}(x)$, now after integrating out the wavefunction in the radial directions, $\psi(x)=\int\mathrm{d}y\,\mathrm{d}z\;\Psi(\textbf{r})$, for $\tau_Q=\{10,60,161\}$, respectively. Panel (d) shows the best fit $\mathcal{X}$ from Eq.\ \eqref{eq:g2GC}, giving an approximate scaling of $\mathcal{X}\sim\hat{\xi}\sim\tau_Q^{0.40(16)}$.
	
	\section{System length dependence}
	\label{app:SMLong}

    In this section, we present results for the freeze-out time and correlation functions in our reduced theory for system sizes consisting of 1024 unit cells, compared to 128 of the main text. The aim of this section is to verify that finite-size effects do not strongly affect the main results of the paper. Due to the large size of the system, we only perform 10 quenches for each $\tau_Q$. 
	
    \begin{figure}[h]
		\centering
		\includegraphics[width=0.9\linewidth]{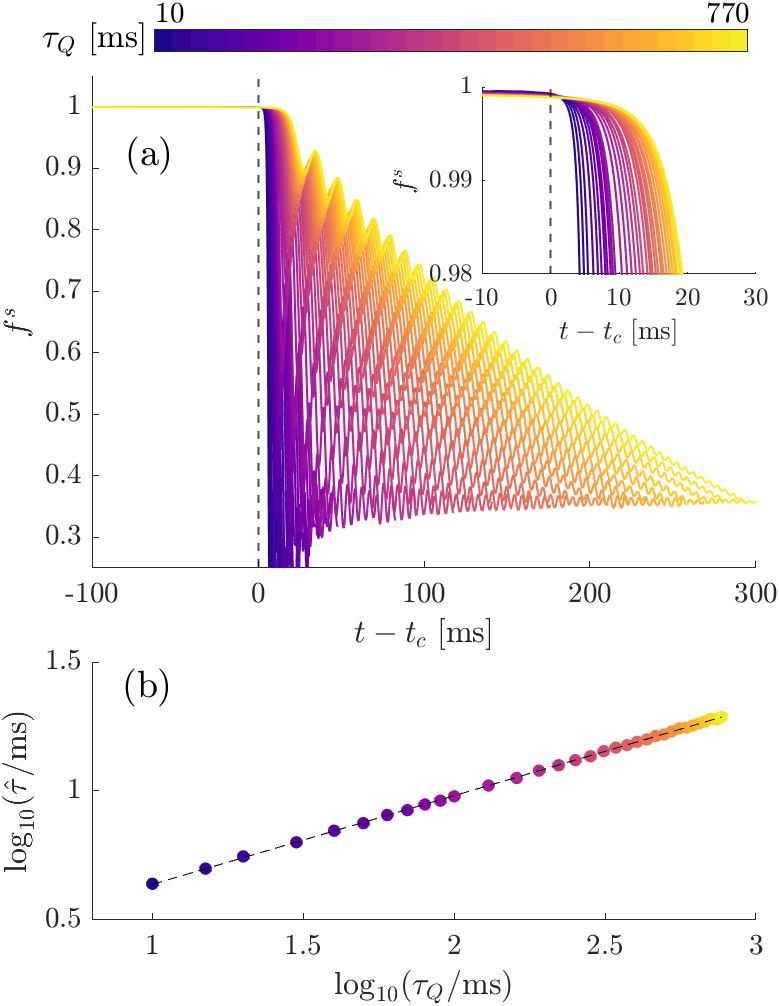}
		\caption{Delay in supersolid formation in the reduced theory for a much longer system size of 1024 unit cells. Similar to Fig.~\ref{fig:freezeout} of the main text, panel (a) shows the full $f^s(t)$, along with a zoomed-in inset. Panel (b) gives the log-log time delay at which $f^s$ crosses 0.98, with the best fit as a dashed line.}
		\label{fig:tDelaylong}
	\end{figure}

    Fig.~ \ref{fig:tDelaylong} shows the superfluid fraction over time in the longer system, governed by Eq.~ \eqref{eq:1DGPE}. The results for the freeze-out time appear to be similar to the results of the shorter system in the main text, with an extracted scaling of $\hat{\tau}_{\text{long}}\propto \tau_Q^{0.344(1)}$. This result is not unexpected: for a uniform system, the delay in supersolid formation should not depend on the system size. It is noteable that relatively few trials are sufficient to give similar results to the main text.

    \begin{figure}[h]
		\centering
		\includegraphics[width=0.9\linewidth]{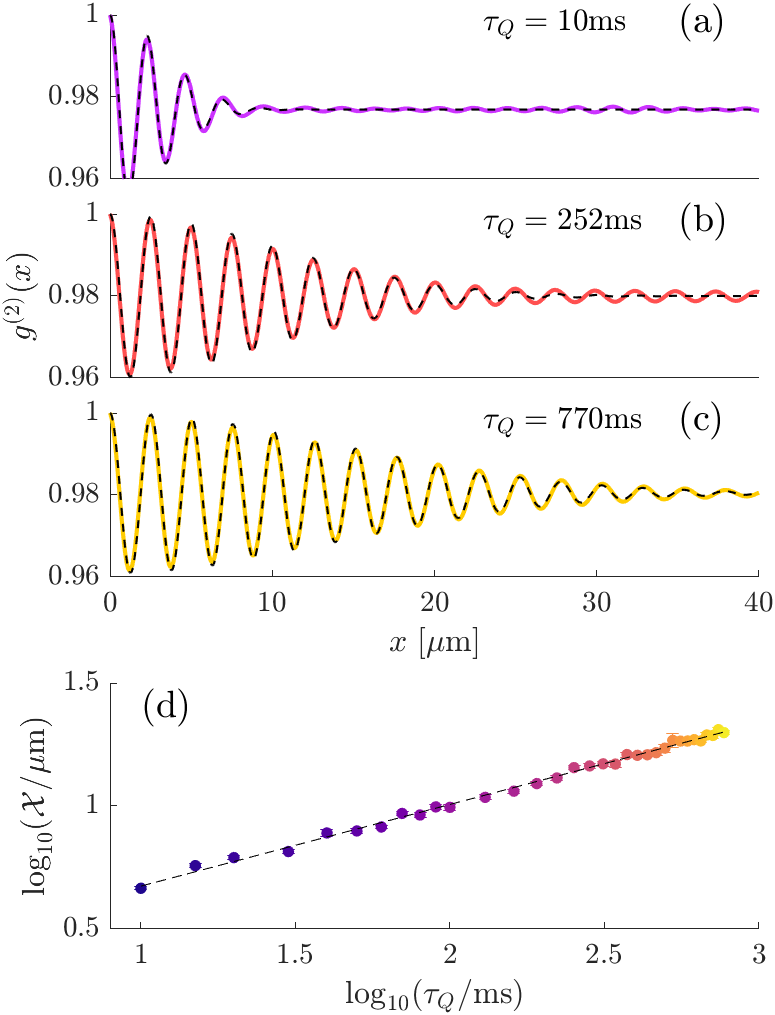}
		\caption{Supersolid correlation length $g^{(2)}(x)$, similar to Fig.~\ref{fig:xi_GC}, now for long system sizes of 1024 unit cells, at $\tau_Q=\{10,252,770\}$ in panels (a)-(c), respectively. The best fit of $\mathcal{X}$ using Eq.~ \eqref{eq:g2GC} is shown as a dashed line.}
		\label{fig:g2long}
	\end{figure}

    In Fig.~\ref{fig:g2long} we show the correlation function $g^{(2)}(x)$ similarly to Fig.~\ref{fig:xi_GC} of the main text, now for the longer system. We again perform a fit according to Eq.~\eqref{eq:g2GC} at the onset of supersolid formation, and extract a scaling result of $\mathcal{X}\sim\hat{\xi}\sim\tau_Q^{0.332(8)}$, similar to the main text. Due to the fact that $g^{(2)}(x)$ is spatially averaged over a much longer system, the results presented here do not suffer from any dramatic loss of statistics due to averaging over only 10 quenches. However, the limited number of quenches does prevent us from achieving statistically meaningful results in terms of defect counting, so we do not attempt any analysis of defect density scaling in the longer system.
    
     
	\section{Relationship between $g^{(2)}$ and $C(x)$}
	\label{app:SMg2Cz}
	
	Assuming the following ansatz for the density,
	\begin{equation}
		n(x)=\bar{n}+\lambda \cos(k x+\phi(x))\;,
	\end{equation}
	where $\phi(x)$ is some locally varying phase, and $\lambda$ is the strength of the supersolid modulation, the correlation function can be determined from
	\begin{align}
		n(x)n(x')
		 =&\;\left[\bar{n}+\lambda \cos(k x+\phi(x))\right]
		\nonumber\\
		&\ \ \times\left[\bar{n}+\lambda \cos(k x'+\phi(x'))\right]
		\nonumber\\
		=&\;\bar{n}^2+\bar{n}\lambda\left[\cos(k x+\phi(x))+\cos(k x'+\phi(x'))\right]\nonumber\\&+\lambda^2\cos(k x+\phi(x))\cos(k x'+\phi(x'))
		\nonumber
        \\
		=&\;\bar{n}^2+\bar{n}\lambda\left[\cos(k x+\phi(x))+\cos(k x'+\phi(x'))\right]
		\nonumber\\
		&+\frac{\lambda^2}{4}\left[\mathrm{e}^{\mathrm{i}(k x+\phi(x))}+\mathrm{e}^{-\mathrm{i}(k x+\phi(x))}\right]
		\nonumber\\
		&\qquad \times\left[\mathrm{e}^{\mathrm{i}(k x'+\phi(x'))}+\mathrm{e}^{-\mathrm{i}(k x'+\phi(x'))}\right]\,.
	\end{align}
	Averaged over realizations, terms like $\cos(k x+\phi(x))$ will vanish, leaving
    \begin{align}
		\langle n(x)n(x')\rangle
		=\bar{n}^2+\frac{\lambda^2}{4}
		&\;\bigl\langle\mathrm{e}^{\mathrm{i}(k (x+x')+\phi(x)+\phi(x'))}\nonumber\\
		&+\mathrm{e}^{\mathrm{i}(k (x-x')+\phi(x)-\phi(x'))}+\text{h.c.}\bigr\rangle
		\nonumber\\
		=\bar{n}^2+\frac{\lambda^2}{4}&\;\bigl\langle\mathrm{e}^{2\mathrm{i}kR}\mathrm{e}^{\mathrm{i}(\phi(x)+\phi(x'))}\nonumber\\
		&+\mathrm{e}^{\mathrm{i}(k s+\phi(x)-\phi(x'))}+\text{h.c.}\bigr\rangle\;,
	\end{align}
	where we let $s=x-x'$ and $R=(x+x')/2$. Now, let us assume that across different trials, the supersolid wavelength is roughly the same, in other words $\langle k\rangle = \bar{k}$, $\langle k^2\rangle = \bar{k}^2$ etc., but variations in the crystal phase $\phi(x)$ is responsible for defects in the coherent crystal. In that case, we can write,
	\begin{align}
		\langle n(x)n(x')\rangle=&\;\bar{n}^2+\frac{\lambda^2}{4}\Biggl[\mathrm{e}^{2\mathrm{i}\bar{k}R}\bigl\langle\mathrm{e}^{\mathrm{i}(\phi(x)+\phi(x'))}\bigr\rangle\nonumber\\
		&+\mathrm{e}^{\mathrm{i}\bar{k} s}\bigl\langle\mathrm{e}^{\mathrm{i}(\phi(x)-\phi(x'))}\bigr\rangle+\text{h.c.}\Biggr]\;,
	\end{align}
	since the exponential prefactors will not change from trial to trial.
	\begin{align}
		\langle n(x)n(x')\rangle=&\;\bar{n}^2+\frac{\lambda^2}{4}\Biggl[\mathrm{e}^{2\mathrm{i}\bar{k}R}\bigl\langle\mathrm{e}^{\mathrm{i}(\phi(x)+\phi(x'))}\bigr\rangle+\text{h.c.}\nonumber\\
		&+\mathrm{e}^{\mathrm{i}\bar{k} s}C^*(x-x')+\mathrm{e}^{-\mathrm{i}\bar{k} s}C(x-x')\Biggr]\;,\\
		\approx&\;\bar{n}^2+\frac{\lambda^2}{2}\cos(\bar{k}s)C(s)
	\end{align}
	where in the last line we have made the following two assumptions: (i) terms which depend on the absolute position of the phase in space, i.e.~$\langle\mathrm{e}^{\mathrm{i}(\phi(x)+\phi(x'))}\rangle$, will average out over many trials; (ii) $C(s)=C(x-x')$ is a real-valued function that only depends on the distance $|x-x'|$. In this way, the decay of  $g^{(2)}(x)$ and $C(x-x')$ appear to be related since they take a similar functional form, see Eq.~\eqref{eq:g2GC}. We note that when extracting $\hat{\xi}$ from $C(x)$ in Fig.~\ref{fig:xi_fwhm}, we do not require $C(x-x')$ to necessarily be exponential (or Gaussian, etc.) -- this is difficult to characterize due to the discrete nature of $C(x-x')$ -- we are simply extracting the width at half maximum.
	
	\section{Universal scaling functions}

	\label{app:SMUniversal}
	\begin{figure}[t]
		\centering
		\includegraphics[width=0.9\linewidth]{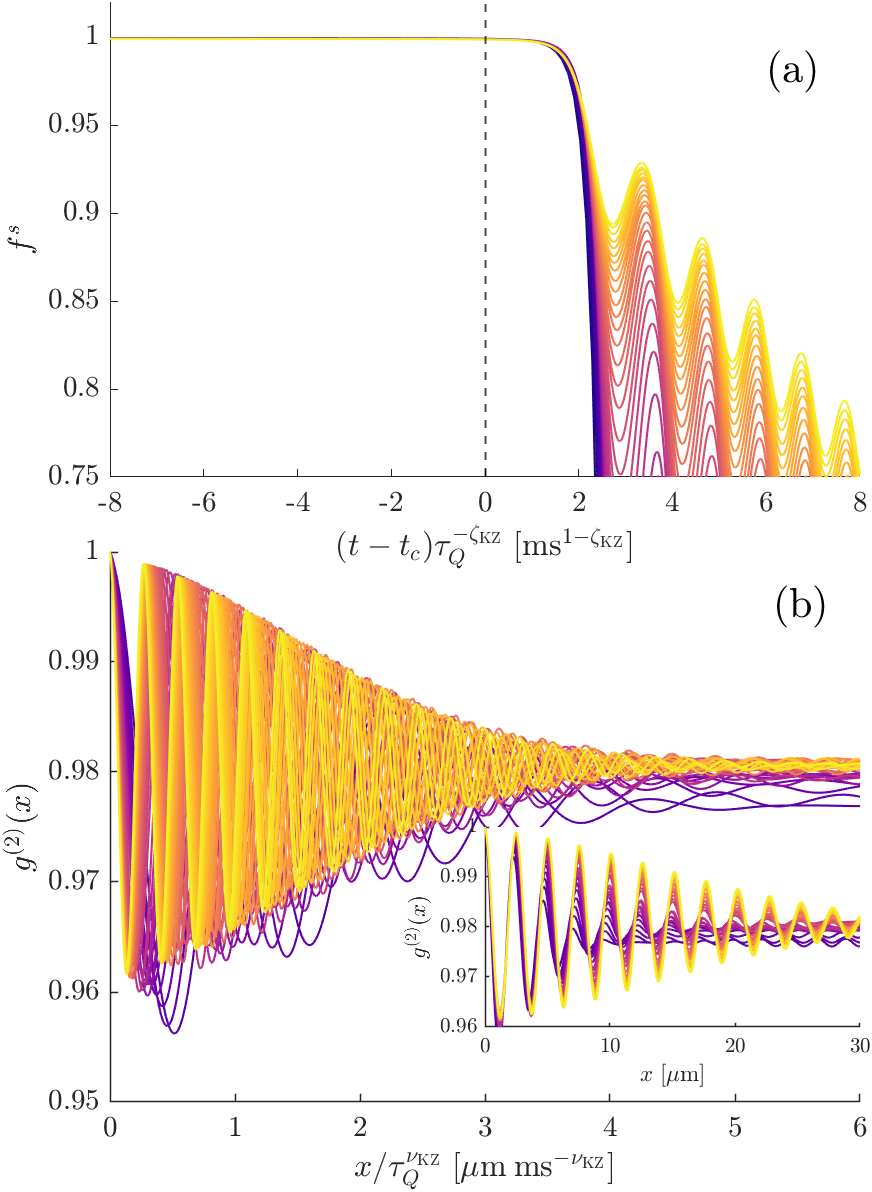}
		\caption{Universal scaling functions. Panel (a) shows the same superfluid fraction data as in Fig.~\ref{fig:freezeout}, now with the time domain rescaled by the extracted Kibble-Zurek exponent $\zeta_{\mathrm{KZ}}$, while (b) shows the data from Fig.~\ref{fig:xi_GC} with the spatial domain rescaled by the extracted $\nu_{\mathrm{KZ}}$. While in (a) the full curves fall on top of one another, in (b) it is only the modulation envelopes that collapse to single curve. The inset shows $g^{(2)}$ over the unscaled domain, demonstrating that the supersolid periodicity is relatively constant across different quench rates.}
		\label{fig:UnivFun}
	\end{figure}
	
	In Fig.~\ref{fig:UnivFun}, we demonstrate that after determining the universal scaling exponents associated with the KZM, the observables can be collapsed to a single universal Kibble-Zurek scaling function. In particular, (a) shows that a rescaling of time with respect to $\zeta_{\mathrm{KZ}}$ allows for the superfluid fraction curves to lie on top of one another, at least within the relatively early regimes we consider to be relevant for the KZM. Later, as more nonlinearities become relevant to the dynamics, there is a departure from the universal scaling function. In (b) there is a similar effect for the $g^{(1)}$ correlator with respect to the exponent $\nu_{\mathrm{KZ}}$ now plotted for all quench rates. It can be seen that rescaling the spatial domain shows that the supersolid oscillation \textit{envelopes} scale in a universal way, but not the oscillations themselves, since the periodicity of the latter is instead set by the roton. Extremely fast quenches ($\tau_Q\lesssim 10\mathrm{ms}$) depart slightly from the universal envelope, likely due to large fluctuations. See the inset of (b) for the unscaled $g^{(2)}$, where it is clear that the oscillations lie on top of one another, while the envelopes do not. Faster quenches also seem to result in a suppression in overall long-range correlations, likely due to the relatively large amount of excitations produced in the system during the quench.

	%

\end{document}